\documentstyle[12pt,epsfig]{article}
\begin{document}

\newcommand{\Title}[1]{{\baselineskip=26pt \begin{center}
            \Large   \bf #1 \\ \ \\ \end{center}}}
\newcommand{\Author}{\begin{center}\large
Gang Rong \\
\vspace{2mm}
Institute of High Energy Physics, Beijing, China\\
\vspace{6mm}
{\em Presented at the 5th International Workshop on Charm Physics, \\
Honolulu, Hawaii, May 14--17, 2012}
\end{center}}
\baselineskip=16pt

\Title{
Review and Report on Results of Leptonic Decays of $D^+$ and $D^+_s$ Mesons
}
\Author

\vspace{0.3cm}

\begin{abstract}

In the last 25 years, many $e^+e^-$ experiments and fixed-target experiments 
performed to search for and study the leptonic decays of the  $D^+$ and $D^+_s$ mesons. 
By 2012, more than 530 signal events of the $D^+$ leptonic decays
and about $4\times10^3$ signal events of the $D^+_s$ leptonic decays 
have been accumulated at these experiments. With these leptonic decay signal events,  
both decay constants $f_{D^+}$ and $f_{D_s^+}$ 
are, respectively, measured at an accuracy level of 
$2.4\%$ and $1.6\%$,
which can be used to more precisely test 
the LQCD calculations of the decay constants.
Comparing these precisely measured $f_{D^+}$ and $f_{D_s^+}$ with those
predicted with theories based on QCD provides some information about New Physics beyond the Standard Model.
In addition to these, with the measured branching fractions for $D^+\rightarrow l^+\nu$
and $D_s^+\rightarrow l^+\nu$ decays, the 
CKM matrix elements 
$|V_{\rm cd}|$ and $|V_{\rm cs}|$ can be determined. 
Comparing these $|V_{\rm cd}|$ and $|V_{\rm cs}|$ to those determined from the CKMfitter
or extracted from $D$ meson semileptonic decays can also provide some information about the New Physics. 
In this article, we review and report the results on the leptonic decays of $D^+$ and $D^+_s$ mesons measured
at different experiments. For the results which have already been published, we review these in shorter
summaries, while for the results which have not been published or have not been reported yet
before we report these with more detailed discussion.  

\end{abstract}

\vspace{1cm}
\section{Introduction}

\begin{figure}
\centerline{
\includegraphics[width=7.5cm,height=3.0cm]{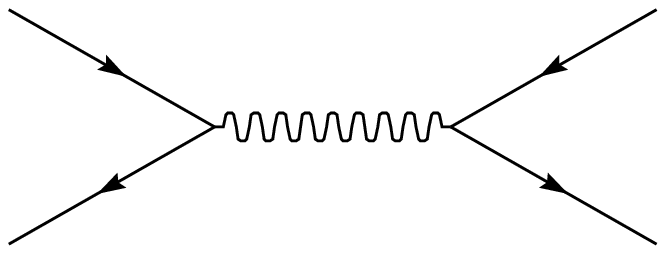}
\put(-265,40){\large\bf $D^+(D_s^+)$}
\put(-15,72){\large\bf $l^+$}
\put(-15,6){\large\bf $\nu_l$}
\put(-214,73){\large\bf $c$}
\put(-214,1){\large\bf $\bar d(\bar s)$}
\put(-115,55){\large\bf $W^+$}
           }
\caption{The decay diagram for $D_{(s)}^+ \rightarrow l^+\nu_l$.}
\label{feynman_digram}
\end{figure}

   In the SM (Standard Model) of particle physics, the $D_{(s)}^+$
(through this article, charge conjugation is implied)
meson can decay into $l^+\nu_l$ (where $l$ is $e$, $\mu$ or $\tau$)
through a virtual $W^+$ boson as shown in Fig.~\ref{feynman_digram}.
The decay rate
is determined by the wavefunction overlap of
the two quarks at the origin, and is parameterized by the $D_{(s)}^+$ decay
constant, $f_{D_{(s)}^+}$.
All strong interaction effects between the two quarks in initial state
are absorbed into $f_{D_{(s)}^+}$.
To the lowest order, as the analogue of the decay width of $\pi^+ \rightarrow l^+\nu_l$, 
the decay width of $D_{(s)}^+ \rightarrow l^+\nu_l$
is given by~\cite{quarks_and_leptons}
\begin{equation}
\Gamma(D_{(s)}^+ \rightarrow l^+\nu_{l})=
     \frac{G^2_F f^2_{D_{(s)}^+}} {8\pi}
      \mid V_{\rm cd(s)} \mid^2
      m^2_l m_{D_{(s)}^+}
    \left (1- \frac{m^2_l}{m^2_{D_{(s)}^+}}\right )^2,
\label{eq01}
\end{equation}
\noindent
\hspace{-0.11cm}where $G_F$ is the Fermi coupling constant, $V_{\rm cd(s)}$ is the
Cabibbo-Kobayashi-Maskawa (CKM) matrix element between the two quarks $c\bar d (\bar s)$~\cite{pdg2010}
in $D^+(D^+_s)$, 
$m_l$ is the mass of the lepton, and
$m_{D_{(s)}^+}$ is the $D_{(s)}^+$ mass.
For this pseudoscalar charged particle leptonic decay, 
the final state neutrino must be left-hand.
Due to angular momentum conservation, the final
lepton must also be left-hand,  
since only in this way one obtain a final state with zero angular momentum component 
in the direction of motion of the leptons. This requirement results in that
the decay rate is proportional to $m_l^2$.
In the limit of $m_l=0$, the $D^+$ and $D^+_s$ leptonic decays are forbidden.
The leptonic decays 
can only occur for the case of $m_l \ne 0$. 
The helicity suppresion of the decay gives a largest decay rate for the final state 
with the lepton $l=\tau$ and gives a larger decay rate for the final state with the lepton $l=\mu$ 
than the one for the final state with lepton $l=e$. 
From the SM, the expected ratios of the decay rates
for $\Gamma(D^+ \rightarrow \tau^+\nu_{\tau})$:$\Gamma(D^+ \rightarrow \mu^+\nu_{\mu})$:$\Gamma(D^+ \rightarrow e^+\nu_e)$
are $2.67:1:2.4\times 10^{-5}$,
while the expected ratios of the decay rates for
$\Gamma(D_s^+ \rightarrow \tau^+\nu_{\tau})$:$\Gamma(D_s^+ \rightarrow \mu^+\nu_{\mu})$:$\Gamma(D_s^+ \rightarrow e^+\nu_e)$
are $9.8:1:2.4\times 10^{-5}$.

In addition to the lowest order decay process, there are some other processes
which increase the $D^+$($D^+_s$) leptonic decay rate. These are radiative decay
and transition to a virtual $D^{*+}$($D^{*+}_s$) by emitting a photon,
where $D^{*+}$($D^{*+}_s$) decays into $l^+\nu$. The latter transition and decay process
is in absence of helicity suppression. These effects should be considered in comparison
of the measured decay rates and the expected decay rates with theories based on QCD.

The pseudoscalar decay constants $f_{D^+}$ and $f_{D_s^+}$ are very important constants 
in heavy flavor physics, 
which connect to $B^+$ and $B^0_s$ mesons decay constants 
$f_{B^+}$ and $f_{B^0_s}$.
As these decay constants are related to the probability of annihilation of the heavy and the light
quarks inside the meson, they play an important role both in characterizing the properties 
of confinement and as absolute normalizations of numerous heavy-flavor transition, including
semileptonic decays and non-leptonic decays of the mesons 
as well as mixing of neutral and anti-neutral meson pairs.
For example, the decay constant 
$f_{B^+(B^0_s)}$ relates to the CKM matrix element $|V_{\rm td(s)}|$ which can be extracted from
the $B^0\bar B^0$ ($B^0_s \bar B^0_s$) mixing experiment. 
However, it is currently not possible to precisely measure $f_{B^+}$
from $B^+$ leptonic decays and is never possible to measure $f_{B^0_s}$ 
since $B^0_s$ does not have leptonic decay,
so theoretical calculations of $f_{B^+}$ and $f_{B^0_s}$
have to be used in determination of $|V_{\rm td}|$ and $|V_{\rm ts}|$.
The decay constants $f_{D^+}$ and $f_{D_s^+}$ as well as $f_{B^+}$ and $f_{B^0_s}$
have been estimated using various theoretical approaches, 
such as QCD-inspired potential model~\cite{QCD-inspired_potential_model},
QCD sum rules~\cite{QCD_sum_rules}, lattice QCD~\cite{simulation_of_QCD_on_lattice},
and alternative non-perturbative methods~\cite{non-perturbative_methods}.
The lattice QCD (LQCD) gives most promising calculations of these decay constants.
The LQCD calculations of the ratios of $f_{D^+}/f_{B^+}$ and $f_{D_s^+}/f_{B^0_s}$ 
are with higher precision
than the calculations of $f_{D^+}$ and $f_{B^+}$ as well as $f_{D_s^+}$ and $f_{B^0_s}$. 
For this reason, we can use  precisely measured $f_{D^+}$ and $f_{D_s^+}$
to valid the LQCD calculations of $f_{D^+}$ and $f_{D^+_s}$.
If the LQCD calculations of $f_{D^+}$ and $f_{D_s^+}$ pass the test 
with the measured $f_{D^+}$ and $f_{D_s^+}$, 
one can use the calculated ratios of $f_{D^+}/f_{B^+}$ and $f_{D_s^+}/f_{B^0_s}$
combined with the precisely measured $f_{D^+}$ and $f_{D_s^+}$ 
to obtain $f_{B^+}$ and $f_{B^0_s}$ with high precision
or one can use the calculated $f_{B^+}$ and $f_{B^0_s}$ with more confidence
to precisely determine
the $|V_{\rm td}|$ and $|V_{\rm ts}|$ in $B^0\bar B^0$ and $B^0_s \bar B^0_s$ 
mixing experiment, respectively.
In addition, with the accurately calculated $f_{B^+}$ 
one can precisely determine the CKM matrix element $|V_{\rm ub}|$.
These improved determinations of $|V_{\rm ub}|$, $|V_{\rm td}|$ and $|V_{\rm ts}|$
would lead to very stringent constraint on the unitary triangle of the CKM matrix.

The CKM matrix elements of $|V_{\rm cd}|$ and $|V_{\rm cs}|$ connect to the
leptonic decays of the $D^+$ and $D^+_s$ mesons.
Historically, measurements of $|V_{\rm cd}|$ were often made
based on the measured branching fractions for $D$ meson semileptonic decays
and inputs of the form factors for these $D$ meson semileptonic decays
or based on the measured neutrino and anti-neutrino interaction.
However, due to largely theoretical uncertainties in calculations of the form factor for
$D\rightarrow \pi l^+\nu_l$ semilepronic decays,
the extracted $|V_{\rm cd}|$ from the measured semileptonic decay branching fractions
suffers from an uncertainty as large as $11\%$~\cite{pdg2010}
and the uncertainty of $|V_{\rm cd}|$ measured from the neutrino and anti-neutrino interaction
is as large as $4.8\%$~\cite{pdg2010} to date.
However, in recent years, the unquenched LQCD calculations of
$f_{D^+}$ have reached a high precision of $\sim2\%$~\cite{lqcd_HPQCD_UKQCD}.
With the precisely
measured branching fraction for $D^+ \rightarrow \mu^+ \nu_{\mu}$
decay together with this precisely
calculated $f_{D^+}$, one can more precisely extract
$|V_{\rm cd}|$.
Similarly, with measured branching fraction of $D^+_s$ leptonic decay,
one can also extract the $|V_{\rm cs}|$.  

There could be some possible new physics effects which contribute to the leptonic decays of the $D^+$ and $D^+_s$
mesons. Dobrescu and Kronfeld~\cite{Dobrescu_and_Kronfeld}, Kundu and Nandi~\cite{Kundu_and_Nandi}
proposed that some non-SM objects participating virtually in the
leptonic decays would modify the decay rates observed experimentally.
To search for the new physics effects, one needs to carefully compare the measured ratio of $f_{D^+_s}/f_{D^+}$
to the one expected with theories based on QCD. Further more,
comparing the values of $|V_{\rm cd}|_{D^+\rightarrow l^+\nu}$ and $|V_{\rm cs}|_{D_s^+\rightarrow l^+\nu}$
extracted from the leptonic decays of $D^+$ and $D_s^+$ mesons to these values of
$|V_{\rm cd}|_{\rm CKMfitter}$ and $|V_{\rm cs}|_{\rm CKMfitter}$ determined from the CKMfitter
or to the values of $|V_{\rm cd}|_{\rm D~semileptonic~decay}$ 
and $|V_{\rm cs}|_{\rm D~semileptonic~decay}$
extracted from $D$ meson semileptonic decays would also provide important information about new physics effects
involved in these leptonic decays.
  
\section{Experiments and Methods}

   The $D^+$ and $D_s^+$ mesons can be produced in different kinds 
of experimental environments,
such as $e^+e^-$ annihilation; 
interaction of hadrons with nuclear targets;
interaction of photons or neutrinos with nuclear targets;
collision of hadrons. In practice, since the ratio of the signal to the background
for leptonic decays of $D^+$ and $D_s^+$ mesons is higher 
at both the $e^+e^-$ collision experiments and fixed target experiments
than that at the hadrons collision experiments, 
all studies of these leptonic decays are performed at $e^+e^-$ collision experiments and
at fixed target experiments.

\subsection{$e^+e^-$ collision near threshold}

    The most clearly experimental environment for studies of these leptonic decays
is the $e^+e^-$ experiments operated near open-charm meson pair production energy
thresholds.
For $D^+$ leptonic decays, the best center-of-mass energy of the $e^+e^-$ collision
is near 3.773 GeV, where the $D^+D^-$ meson pairs are produced. 
Searching for the leptonic decay of $D^+$ meson and 
measurements of leptonic decay branching fractions and decay constant
of $D^+$ meson were made at the historical detectors of MARK-III, 
BES-I, BES-II and CLEO-c, and today's running BES-III.

    Taking the advantage of the $D^+D^-$ production, one first accumulates the
samples of the reconstructed $D^-$ mesons in one side, then can absolutely
measure the branching fraction for $D^+$ leptonic decay
by examining the decay products in the
system recoiling against the $D^-$ tags. In the data analysis, one can search
for the $D^+\rightarrow l^+\nu$ in the recoil of the singly tagged $D^-$ mesons 
by calculating the missing mass square, which is the missing energy square minus the missing momentum square.
If there is a neutrino in the recoil side of the tagged $D^-$ meson, the
distribution of the missing mass square should characterize with a peak at
zero. By examining this missing mass square distribution of the singly tagged $D^-$ mesons
together with one charged track which is identified as a lepton, one can fully reconstruct
the leptonic decay of $D^+\rightarrow l^+\nu$. Based on the numbers of the fully reconstructed
$D^+\rightarrow l^+\nu$ events and the singly tagged $D^-$ mesons, one can well measure
the branching fraction for $D^+\rightarrow l^+\nu$ decays, and determine the decay constant
$f_{D^+}$. 

   Similarly, for $D_s^+$ leptonic decays, the best center-of-mass energy 
of the $e^+e^-$ collision is near 4.03 (or 4.17) GeV, 
where the $D_s^+D_s^-$ ($D_s^+D_s^{*-}$) meson pairs are produced. 
Historically, the BES-I experiment accumulated data at 4.03 GeV and 4.14 GeV, 
while the CLEO-c accumulated data at 4.17 GeV, and BES-III accumulated data at 4.01 GeV.
The method of measurement of $D_s^+$ leptonic decay branching fraction and decay constant
at these energies are almost the same as
these for measurement of $D^+$ leptonic decay branching fraction and decay constant at 3.773 GeV.

\subsection{$e^+e^-$ collision at higher energy}

   Since the $D^+$ and $D_s^+$ mesons can be formed in quark fragmentation,
in principle these leptonic decay branching fractions can also be measured 
by analyzing the data taken at the $e^+e^-$ experiments
operated near 10.5 GeV and 91 GeV, where the $B$ factory experiments and
$Z^0$ physics experiments were performed.
However, since the $D^+$ leptonic decays are Cabibbo-suppressed decays,
it is difficult to measure the leptonic decay branching fractions and decay constant of $D^+$ meson
with the data collected at these two energies. The data taken at these two energies can be used to measure
the leptonic decay branching fractions and decay constant of $D_s^+$ meson.

   The CLEO experiment at the CESR storage ring, 
BaBar experiment at PEP-II,
and BELLE experiment at asymmetric-energy collider (KEKB) 
collected or have been collecting large data samples near 10.5 GaV.
While ALEPH, L3, and OPAL experiments at the LEP 
accumulated large data samples of $Z^0$ hadronic decay events at 91 GeV.
With these large data samples of $e^+e^- \rightarrow c\bar c$ events and 
$e^+e^- \rightarrow Z^0 \rightarrow c\bar c$ events, 
the CLEO, BaBar, BELLE, ALEPH, L3, and OPAL experiments measured the branching fractions for
$D_s^+ \rightarrow l^+\nu$ decays and determined the decay constant $f_{D_s^+}$.

    The analysis method used in measurements of the branching fractions 
for $D_s^+ \rightarrow l^+\nu$ with the data taken at 10.5 GeV
required the unfolding of the fragmentation process. The total number of
$D_s^-$ mesons in the data sample is estimated by reconstructing the four momentum of $D_s^-$ 
candidates recoiling against the rest of the events. The number of $D_s^+ \rightarrow l^+\nu$
events is obtained by identifying a lepton candidate and reconstructing the whole event, 
including the missing neutrino system.

    The analysis method used by ALEPH, L3 and OPAL experiments are based on analysis of 
fragmentation and decay chain $Z^0\to c\bar c, ~c\to D^{*+}_s$ followed by 
$D^{*+}_s\to \gamma D_s^+,~D^+_s\to\tau^+\nu_{\tau}, ~\tau^+\to l^+\nu_e\nu_l$.

\subsection{Fixed-target experiments}

     The fixed-target experiment is other kind of experiment at which the charm mesons
can be produced in the interaction of the incident particles with nucleus of the target.
From these daughter particles coming from the interaction the $D^+$ and $D_s^+$ mesons
can be selected and their decay properties can be studied.
The cross sections of charm meson production in fixed-target experiments
are higher than these at the $e^+e^-$ experiments. However, the non-charm background in the
fixed-target experiment are much higher than these at the $e^+e^-$ experiments. 
To reduce the background events for studying the charm meson decays, the decay length of events are often
measured with the vertex detector which is placed near the target. 
Since the charm mesons have relative long lifetimes, 
they can travel measurable distances from primary production point. 
Using the technique of reconstruction of the second vertex of the charm mesons decay, 
one can well separate the charm meson decay events from light hadron events.

    Three fixed-target experiments, the CERN WA75~\cite{WA75}, CERN WA92~\cite{WA92} 
and Fermilab E653~\cite{E653} studied the $D_s^+$ leptonic decays.
The WA75~\cite{WA75} experiment is an emulsion-hybrid experiment, 
which is   designed to search for charm quark pair production
in 350 GeV/$c$ $\pi^-$ nucleus interactions. A total of about 80 liters of nuclear emulsion is exposed
to a $\pi^-$ beam from the CERN SPS.
The WA92~\cite{WA92} is designed to study the production and decays of beauty particles 
from 350 GeV/$c$ $\pi^-$ interaction in copper and tungsten. Charged particle tracking is performed using the omega
spectrometer. The charmed meson decays can also be reconstructed with the spectrometer together 
with a silicon vertex detector placed near the target.
The E653~\cite{E653} experiment is also an emulsion-hybrid experiment designed to study production and decays of
heavy flavor particles by the direct observation of decay vertex in the emulsion. The charm mesons
are from the interaction of a 600 GeV/$c$ $\pi^-$ and  nucleus of the target.
These three experiments selected the purely leptonic decays of $D_s^+ \rightarrow l^+\nu$
by using transverse momentum spectrum of muons from
$D_s^+$ leptonic decay observed in an emulsion target.

\section{Leptonic decays of $D^+$ meson}

Several experiments performed to search for the leptonic decays of $D^+$ meson
and to precisely measure its leptonic decay branching fractions
and decay constant $f_{D^+}$ in the last 25 years. In this section, we first review
the available measurements of its leptonic decay branching fractions and decay constant 
which have been already published, then on behalf of the BES-III collaboration we report
new results of precision measurements of the branching fraction for
$D^+ \rightarrow \mu^+\nu$ decays and decay constant $f_{D^+}$ which are obtained at the BES-III experiment.
 
\subsection{Review of results at old experiments}

\subsubsection{Search for $D^+ \rightarrow l^+\nu$ decay at Mark-III experiment}

   In 1988, MARK III collaboration first searched for the decay of $D^+ \rightarrow l^+\nu$. 
The MARK-III did not observe any signal events for this decay. 
They set an upper limit on the decay constant, which is less than 290 MeV at $90\%$ C.L.~\cite{mark-iii_fD}.

\subsubsection{First measurements of $B(D^+ \rightarrow l^+\nu)$ and  $f_{D^+}$ at the BES experiments}

In 1998, the BES collaboration analyzed 22.3 pb$^{-1}$ of data taken at 4.03 GeV. 
From 5 single $D$ tag modes, they found 10082 $D^+$ mesons produced in their data sample. 
From this data sample, they found 1 event for $D^+\rightarrow \mu^+\nu$ decay, 
and measured the branching fraction for $D^+\rightarrow \mu^+\nu$ to be 
$(0.08^{+0.16+0.05}_{-0.05-0.02})\%$, corresponding to a value of 
decay constant of $f_{D^+}=(300^{+180+80}_{-150-40})$ MeV~\cite{bes-i_fD}.

In 2004, the BES collaboration analyzed 33 pb$^{-1}$ of data taken in $e^+e^-$ annihilation with
their upgraded BES-II detector at the BEPC collider to study the leptonic decays of $D^+$ meson.
From 9 single $D^-$ tag modes, they accumulated $5321\pm 149\pm 160$ $D^-$ tags. 
In the system recoiling against the $D^-$ tags, 
they found 
3 signal events for $D^+ \rightarrow \mu^+\nu$ decays with 0.3 background events 
estimated with Monte Carlo simulation or estimated with the same data set. 
With these signal events and the $5321\pm 149\pm 160$ $D^-$ tags, 
they measured the branching fraction for $D^+ \rightarrow \mu^+\nu$ decays to be 
$B(D^+ \rightarrow \mu^+\nu)=(0.122^{+0.111}_{-0.053}\pm 0.010)\%$, corresponding to
a value of the decay constant $f_{D^+}=(371^{+129}_{-119}\pm 25)$ MeV~\cite{bes-ii_fD}. 
These are absolute measurements of the decay branching fraction and decay constant, which do
not depend on the yield of $D^+$ meson production and do not depend on some branching fractions
for $D^+$ meson decay into other modes.

\subsubsection{Measurements of $B(D^+ \rightarrow l^+\nu)$ and $f_{D^+}$ at CLEO-c experiment}

In 2004, the CLEO collaboration analyzed 60 pb$^{-1}$ of data taken in $e^+e^-$ annihilation
at 3.770 GeV with the CLEO-c detector at the CESR . From 5 single $D^-$ tag modes, 
they found $28574\pm207\pm629$ $D^-$ tags. In the system recoiling against the $D^-$ tags, they found 7 signal events 
for $D^+ \rightarrow \mu^+\nu$ decays, and measured the branching fraction for 
$D^+ \rightarrow \mu^+\nu$ decays to be $B(D^+ \rightarrow \mu^+\nu)=(3.5\pm 1.4 \pm 0.6)\times 10^{-4}$,
corresponding to a value of the decay constant $f_{D^+}=(202 \pm 41 \pm 17)$ MeV~\cite{cleo-c_fD_2004}.

In 2005, using 281 pb$^{-1}$ of data taken at 3.770 GeV the CLEO collaboration presented
$47.2 \pm 7.1^{+0.3}_{-0.8}$ signal events for $D^+ \rightarrow \mu^+\nu$ decay observed in the system
recoiling against $158354\pm 496$ $D^-$ tags. They measured the decay branching fraction of
$B(D^+ \rightarrow \mu^+\nu)=(4.40 \pm 0.66^{+0.09}_{-0.12})\times 10^{-4}$ and
extracted the decay constant $f_{D^+}=(222.6 \pm 16.7^{+2.8}_{-3.4})$ MeV~\cite{cleo-c_fD_2005}.

In 2008, the CLEO collaboration accumulated $460055\pm 787$ $D^-$ tags with 6 hadronic decay modes
of the $D^-$ meson from all of 818 pb$^{-1}$ of data taken at 3.773 GeV.
They presented $149.7 \pm 12.0$ signal events 
for $D^+ \rightarrow \mu^+\nu$ decays observed in the system recoiling against these $D^-$ tags.
They claimed that they measured the decay branching fraction of 
$B(D^+ \rightarrow \mu^+\nu)=(3.82 \pm 0.32 \pm 0.09)\times 10^{-4}$
and determined the decay constant of $f_{D^+}=(205.8 \pm 8.5 \pm 2.5)$ MeV~\cite{cleo-c_fD_2008}.
In measurement of this decay branching fraction and determination of the decay constant,
the CLEO assumed that the ratio of the number of the signal events 
for $D^+ \rightarrow \mu^+\nu$ decay
over the number of the background events for $D^+ \rightarrow \tau^+\nu$ decay
in their fitted missing mass squared region is a constant, and they fixed this ratio to
the Standard Model value. However, this is not the case of the experimental observation
due to that both the number of the events for $D^+ \rightarrow \mu^+\nu$ decays and
the number of the events for $D^+ \rightarrow \tau^+\nu$ decays fluctuate.
In addition to these, they assumed that the number of background
events do not fluctuate, so they fixed the number of background events in their determination
of the number of the signal events. In this case, they obtained the statistical uncertainty
in the number of net signal events to be  smaller than the square root of the number
of the signal events~\footnote{For example, $\sqrt{149.7}=12.2$ which is larger than 12.0, 
where 12.0 is the CLEO reported error of $149.7$ signal events observed.}. 
In this case, CLEO collaboration reported their measured branching fraction 
and the decay constant as mentioned above. 

However, in the CLEO published paper~\cite{cleo-c_fD_2008}, 
they also gave conservative results of the decay branching fraction and decay constant, 
which are $B(D^+ \rightarrow \mu^+\nu)=(3.93 \pm 0.35 \pm 0.09)\times 10^{-4}$
and $f_{D^+}=(207.6 \pm 9.3 \pm 2.5)$ MeV.
These branching fraction and decay constant were determined in the case of that
both the number of events for $D^+ \rightarrow \mu^+\nu$ decays and
the number of events for $D^+ \rightarrow \tau^+\nu$ decays
were allowed to be fluctuated in their fit.
So these experimental results are more reliable. 
But these are not appeared in neither the {\bf Abstract} or
the {\bf Conclusions} of the CLEO published paper~\cite{cleo-c_fD_2004}.
We will use $B(D^+ \rightarrow \mu^+\nu)=(3.93 \pm 0.35 \pm 0.09)\times 10^{-4}$
and $f_{D^+}=(207.6 \pm 9.3 \pm 2.5)$ MeV for our
further discussion in this article.

In the system recoiling against $460055\pm 787$ $D^-$ tags, the CLEO collaboration found $27.8\pm 16.4$ $\tau^+ \nu$ with
$\tau^+ \rightarrow \pi^+\bar \nu$ events in their missing mass squired range for the signal. They set an upper limit
on the decay branching fraction of $B(D^+\rightarrow \tau^+\nu)<1.2 \times 10^{-3}$ at $90\%$ C.L..

\subsection{New results at BES-III experiment}

With the BES-III detector~\cite{bes3} at the BEPC-II~\cite{bepc2}, 
the BES-III collaboration collected 2.89 fb$^{-1}$ of data at
3.773 GeV during the time period from 2010 to 2011. 
With this data sample, the BES-III made precision measurements of
the decay branching fraction for $D^+\rightarrow \mu^+\nu_{\mu}$
and decay constant $f_{D^+}$.
In this section, we report measurements of
the branching fraction for $D^+\rightarrow \mu^+\nu_{\mu}$ decay and
the pseudoscalar decay constant $f_{D^+}$
obtained by analyzing this data sample.

The singly tagged $D^-$ mesons are reconstructed in nine
non-leptonic decay modes of
$K^+\pi^-\pi^-$,
$K^0_s\pi^-$,
$K^0_s K^-$,
$K^+K^-\pi^-$,
$K^+\pi^-\pi^-\pi^0$,
$\pi^+\pi^-\pi^-$,
$K^0_s\pi^-\pi^0$,
$K^+\pi^-\pi^-\pi^-\pi^+$,
and
$K^0_s\pi^-\pi^-\pi^+$.
Events which contain at least
three reconstructed charged tracks with good helix fits
and their $|{\rm cos\theta} |<0.93$
are selected, where $\theta$ is the polar angle of the charged tracks.
All tracks, save those from $K^0_s$ decays,
must originate from the interaction region,
which require that the closest approach of a charged track
in the $xy$ plane is less than 1.0 cm and
is less than
15.0 cm in the $z$ direction.
Pions and kaons are identified by means of TOF and $dE/dx$ measurements
with which the combined confidence levels $CL_{\pi}$ and $CL_{K}$ for pion and kaon
hypotheses are, respectively, calculated.
Pion (kaon) identification requires
$CL_{\pi}>CL_{K}$ ($CL_{K}>CL_{\pi}$) for its momentum $p<0.75$ GeV/$c$
and $CL_{\pi}>0.1\%$ ($CL_{K}>0.1\%$) for its momentum $p\ge 0.75$ GeV/$c$.

   To select good photons from the $\pi^0$ meson decays,
the energy of the photon deposited in the barrel (end-cap) EMC
is required to be greater than $0.025~(0.050)$ GeV.
The barrel (end-cap) EMC covers the range of
$|\rm {cos \theta_{\gamma} }|<0.83~(0.85\le |\rm {cos \theta_{\gamma} }|<0.93)$,
where $\theta_{\gamma}$ is the polar angle of the photon.
In addition,
the EMC cluster timing TDC is required to be in the range of $0 \leq {\rm TDC} \leq 700$ ns.
In order to reduce background the angle between the
photon and the nearest charged track is required to be greater than $10^{\circ}$.
To further reduce the combinatorial background, the 1-C kinematic fit
is performed to constrain the invariant mass of $\gamma\gamma$ to the mass of $\pi^0$ meson.
If the 1-C kinematic fit is successful
these $\gamma\gamma$
are kept as good candidates for $\pi^0 \rightarrow \gamma\gamma$ decay.

    To select $K^0_s$ decays,
a second vertex fit is subjected to two charged tracks with opposite
charge and the $\chi^2$ from the vertex fit is required
to be less than 999.0.
In addition, the secondary vertex from which the $\pi^+\pi^-$ pair originate should be
displaced from the event vertex at least by the decay length $L_{xyz} >0$ mm. After these, only
the $\pi^+\pi^-$ meson pair with invariant mass $M_{\pi^+\pi^-}$ being within
about $\pm 3.5\sigma$ mass window of the nominal $K^0_s$ mass is taken as the $K^0_s$ meson
candidate.

   The singly tagged $D^-$ mesons are fully reconstructed by requiring
the difference in the energy, $\Delta E$, of the daughter particle
$mKn\pi$ (where m=0, 1, 2; n = 0, 1, 2, 3, or 4) system
with the beam energy.
They then require
$|\Delta E|<(2 \sim 3)\sigma_{E_{mKn\pi}}$,
where $\sigma_{E_{mKn\pi}}$ is the standard deviation
of the distribution of the energy of $mKn\pi$ system,
and then examine the beam energy constraint mass
of the tagged $mKn\pi$ system,
\begin{equation}
M_{\rm B} = \sqrt {E_{\rm beam}^2-|\vec {p}_{mKn\pi}|^2},
\end{equation}
where $E_{\rm beam}$ is the beam energy,
and
$|\vec p_{mKn\pi}|$ is the magnitude of the momentum
of the daughter particle $mKn\pi$ system.

\begin{figure}[htbp]
\centerline{
\includegraphics[width=12.0cm,height=8.0cm]{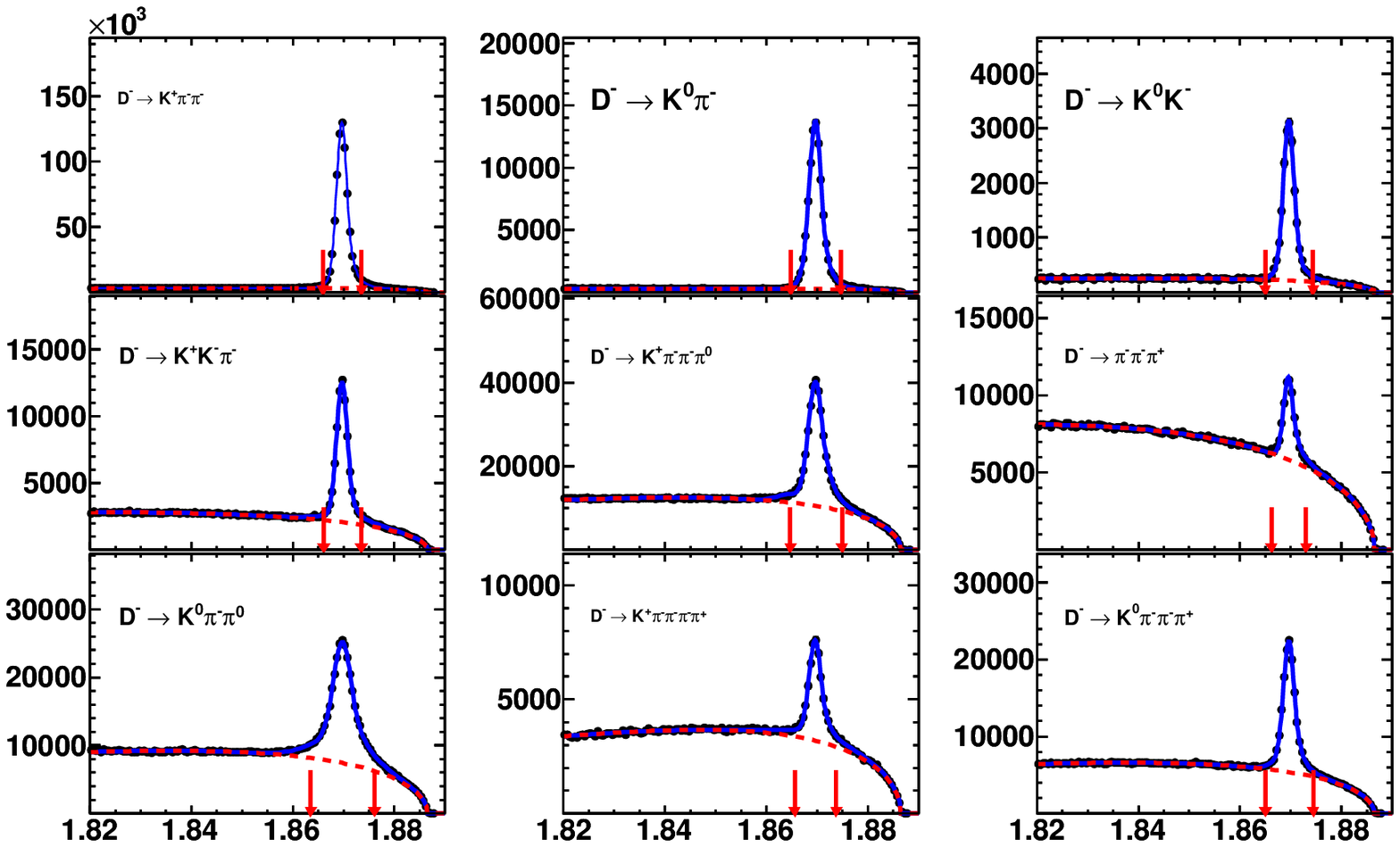}
\put(-212.0,5.0){{M$_{\rm B}$ ~~[GeV/$c^2$]}}
\put(-350.0,80){\rotatebox{90}{Number of Events}}
\put(-300.0,160.0){(a)}
\put(-195.0,160.0){(b)}
\put(-95.0, 160.0){(c)}
\put(-300.0,100.0){(d)}
\put(-195.0,100.0){(e)}
\put(-95.0, 100.0){(f)}
\put(-300.0,50.0){(g)}
\put(-195.0,50.0){(h)}
\put(-95.0, 50.0){(i)}
        }
\caption{Distributions of the beam energy constraint masses
of the $mKn\pi$
combinations for the 9 single tag modes from the data;
where
(a), (b), (c), (d), (e), (f), (g), (h), (i)
are for the modes of $D^- \rightarrow K^+\pi^-\pi^-$,
$D^- \rightarrow K^0_s\pi^-$,
$D^- \rightarrow K^0_s K^-$,
$D^- \rightarrow K^+K^-\pi^-$,
$D^- \rightarrow K^+\pi^-\pi^-\pi^0$,
$D^- \rightarrow \pi^+\pi^-\pi^-$,
$D^- \rightarrow K^0_s\pi^-\pi^0$,
$D^- \rightarrow K^+\pi^-\pi^-\pi^-\pi^+$,
and
$D^- \rightarrow K^0_s\pi^-\pi^-\pi^+$,
respectively.
}
\label{fig3}
\end{figure}

   The $M_{\rm B}$ distributions for the nine $D^-$ tag modes are shown
in Fig.~\ref{fig3}.
A maximum likelihood fit
to the mass spectrum with a Crystal Ball function plus an Gaussian function
for the $D^-$ signal and
the ARGUS function to describe background
yields the number of the singly tagged $D^-$ events
for each of the nine modes.
Selecting these candidates for $D^-$ tags within
the range marked by arrows in Fig.~\ref{fig3}
reduce signal number by about $2\%$ giving a total of $1586056 \pm 2327$ $D^-$ tags.
In these $D^-$ tags,
20103 $D^-$ tags are reconstructed in more than one single $D^-$ tag mode.
Subtracting this number of the double counting $D^-$ tags from the $1586056 \pm 2327$ $D^-$ tags
yields $1565953 \pm 2327$ $D^-$ tags which are used
for further analysis of measuring
the branching fraction for $D^+ \rightarrow \mu^+\nu_{\mu}$ decays.

Candidate events for the decay $D^+ \rightarrow \mu^+\nu_{\mu}$
are selected from the surviving charged tracks in the system recoiling against the
singly tagged $D^-$ mesons.
To select the $D^+ \rightarrow \mu^+ \nu_{\mu}$,
it is required that there be a single charged track
originating from the interaction region
in the system recoiling against the $D^-$ tag and the charged track
satisfies $|\rm {cos \theta}|<0.93$
as well as
it is identified
as a $\mu^+$.
The $\mu^+$ can be well identified with the passage length
of a charged particle through the MUC since a charged hadron (pion or kaon)
quickly loses its energy due to its
strong interactions with the absorber of the MUC
and most of the hadrons stop in the absorber before passing a long passage length in the MUC.
For the candidate event,
no extra good photon which is not used in the reconstruction
of the singly tagged $D^-$ meson
is allowed to be present in the event,
where the ``good photon'' is the one with deposited energy in the EMC being greater than 300 MeV.

Since there is a missing neutrino
in the purely leptonic decay event, the event should
be characteristic with missing energy $E_{miss}$ and
missing momentum $p_{miss}$
which are carried away by the neutrino.
So they infer the existence of the neutrino by requiring a measured
value of the missing mass squared $M^2_{\rm miss}$
to be around zero. The missing mass squared $M^2_{\rm miss}$
is defined as
\begin{equation}
M^2_{\rm miss} = (E_{\rm beam}-E_{\mu^+})^2 - (- \vec p_{D^-_{\rm tag}}- \vec p_{\mu^+} )^2,
\label{eq_miss2}
\end{equation}
where $E_{\mu^+}$ and $\vec p_{\mu^+}$ are, respectively, the energy and three-momentum of the $\mu^+$,
and $\vec p_{D^-_{\rm tag}}$ is three-momentum of the candidate for $D^-$ tag.

\begin{figure}[htbp]
\centerline{
\includegraphics[width=12.0cm,height=11.0cm]{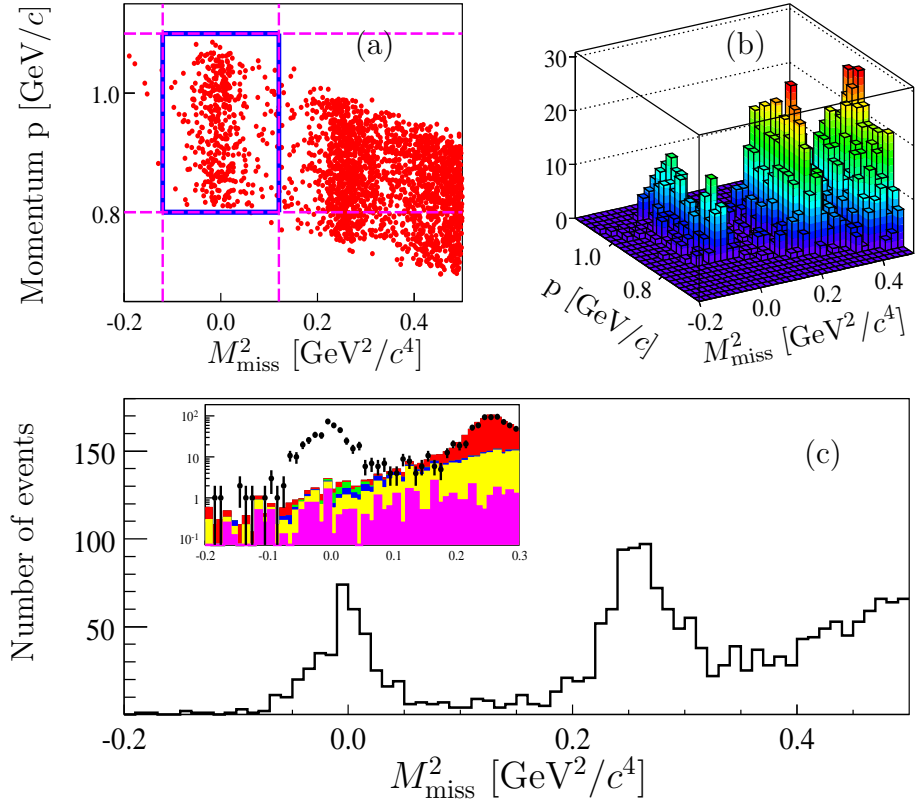}
\put(-276.0,158.0){$M^2_{\rm miss}$~[GeV$^2$/$c^4$]}
\put(-347.0,180){\rotatebox{90}{Momentum p [GeV/$c$]}}
\put(-90.0,158.0){\rotatebox{10}{\small $M^2_{\rm miss}$~[GeV$^2$/$c^4$]}}
\put(-150.0,186){\rotatebox{-32}{\small p [GeV/$c$] }}
\put(-207.0,-3.0){\large{$M^2_{\rm miss}$~[GeV$^2$/$c^4$]}}
\put(-350.0,45){\rotatebox{90}{Number of events}}
\put(-220.0,275.0){(a)}
\put(-80.0,275.0){(b)}
\put(-50.0, 120.0){(c)}
           }
\caption{Distributions of $M^2_{\rm miss}$, where (a) and (b) are scatter plots
of the identified muon momentum $p$ VS $M^2_{\rm miss}$,
and (c) is the distribution of $M^2_{\rm miss}$.
The insert shows the signal region for $D^+ \rightarrow \mu^+\nu_{\mu}$
on a log scale, where dots with error bars are for the data,
histograms are for the simulated backgrounds from
$D^+ \rightarrow K^0_L \pi^+$ (red),
$D^+ \rightarrow \pi^0\pi^+$ (green),
$D^+ \rightarrow \tau^+ \nu_{\tau}$ (blue) and
other decays of $D$ mesons (yellow) as well as
from $e^+e^- \rightarrow$non-$D\bar D$ decays (pink).
}
\label{pmu_vs_umiss_bes3}
\end{figure}

     Figure~\ref{pmu_vs_umiss_bes3}(a) and (b) show the scatter-plots
of the momentum of the identified muon satisfying the requirement
for selecting $D^+\rightarrow \mu^+\nu_{\mu}$ decay versus $M^2_{miss}$,
where the blue box in Fig.~\ref{pmu_vs_umiss_bes3}(a) shows the signal region
for $D^+\rightarrow \mu^+\nu_{\mu}$ decays.
Within the signal region,
there are $425$ candidate events for $D^+ \rightarrow \mu^+\nu_{\mu}$ decay.
The two concentrated clusters out side of the signal region are
from $D^+$ non-leptonic decays and some other background events.
The events whose peak is around 0.25 GeV$^2$/$c^4$ in $M^2_{miss}$ are
mainly from $D^+ \rightarrow K_L^0\pi^+$ decays, where $K_L^0$ is missing.
Projecting the events for which the identified muon momentum being in the range from 0.8 to 1.1 GeV/$c$
onto the horizontal scale yields
the $M^2_{miss}$ distribution as shown in Fig.~\ref{pmu_vs_umiss_bes3}(c),
where the difficultly suppressed backgrounds from
$D^+ \rightarrow K_L^0\pi^+$ decays in CLEO-c measurement~\cite{cleo-c_fD_2008}
are effectively suppressed due to that they use the MUC measurements to identify the muon.

\begin{table}
\caption{Sources of background events for $D^+ \rightarrow \mu^+\nu_{\mu}$.}
\label{Nb_bck}
\centerline{
\begin{tabular}{lr} \hline\hline
Source mode  & Number of events \\
\hline
$D^+\to K^0_L \pi^+$            &                      $7.9 \pm 0.8$ \\
$D^+\to \pi^+ \pi^0$            &                      $3.8 \pm 0.5$ \\
$D^+\to \tau^+ \nu_{\tau}$      &                      $6.9 \pm 0.7$ \\
Other decays of $D$ mesons      &                      $17.9 \pm 1.1$ \\
$e^+e^-\to \gamma\psi(3686)$    &                      $0.2 \pm 0.2$  \\
$e^+e^-\to \gamma J/\psi$       &                      $0.0 \pm 0.0$ \\
$e^+e^-\to light~hadron$ (continuum) &                 $8.2 \pm 1.4$ \\
$e^+e^-\to \tau^+\tau^-$        &                      $1.9 \pm 0.5$ \\
$\psi(3770)\to non-D\bar D$     &                      $ 0.9\pm0 .4$ \\ \hline
Total                           &                      $47.7 \pm 2.3$ \\
\hline \hline
\end{tabular}
      }
\end{table}

    Some non-purely leptonic decay events from the $D^+$, $D^0$,
$\gamma \psi(3686)$, $\gamma J/\psi$,
$\psi(3770) \rightarrow {\rm non}-D\bar D$,
$\tau^+\tau^-$ decays as well as continuum light hadron production
may also satisfy the selection criteria
and are the background events to the purely
leptonic decay events.
These background events must be subtracted off.
The number of the background events can be estimated by analyzing
different kinds of Monte Carlo simulation events.
Detailed Monte Carlo studies show that there are
$47.7 \pm 2.3 \pm 1.3$
background events in $425$ candidates
for $D^+ \rightarrow \mu^+\nu_{\mu}$ decays,
where the first error is due to Monte Carlo statistic and second systematic
arising from uncertainties in the branching fractions 
or production cross sections for the source modes as shown in Table~\ref{Nb_bck}.

    After subtracting the number of background events,
$377.3\pm 20.6 \pm 2.6$ signal events
for $D^+ \rightarrow \mu^+\nu_{\mu}$ decay are retained,
where the first error is statistical and the second systematic arising from the uncertainty
of the background estimation.

\begin{table}
\caption{Sources of the relative systematic uncertainties
in the measured branching fraction for $D^+ \rightarrow \mu^+\nu_{\mu}$ decay.}
\label{src_sys_err}
\begin{center}
\begin{tabular}{lr} \hline\hline
Source  & Systematic uncertainty [$\%$] \\
\hline
Number of $D^-$ tags (N$_{D^-_{tag}}$)                 & 0.6                     \\
Muon tracking                                          & 0.5                     \\
$\mu$ selection                                        & 0.3                     \\
E$_{\gamma_{\rm max}}$ cut                             & 0.7                     \\
Muon momentum cut                                      & 0.1                     \\
$M^2_{miss}$ cut                                       & 0.5                     \\
Background estimation                                  & 0.7                     \\
Monte Carlo statistics                                 & 0.2                     \\
Radiative correction                                   & 1.0                     \\
\hline
Total                                                  & 1.7                     \\
\hline \hline
\end{tabular}
\end{center}
\end{table}

    The overall efficiency for observing the decay ${D^+\rightarrow \mu^+\nu_{\mu}}$
is obtained by analyzing full Monte Carlo simulation events
of $D^+\rightarrow \mu^+\nu_{\mu}$ VS $D^-$ tags
and combining with $\mu^+$ reconstruction efficiency in the MUC.
The $\mu^+$ reconstruction efficiency in the MUC is measured with
muon samples selected from the same data taken at 3.773 GeV. The overall efficiency is
$0.6382~\pm 0.0015$.

   With $1565953$ singly tagged $D^-$ mesons, $377.3\pm 20.6 \pm 2.6$
$D^+ \rightarrow \mu^+\nu_{\mu}$ decay events observed and
the efficiency $0.6382~\pm 0.0015$,
the BES-III collaboration obtain the branching fraction
$$B(D^+ \to \mu^+\nu_{\mu})=(3.74 \pm 0.21 \pm 0.06)\times 10^{-4}~~({\rm BESIII~Preliminary}), $$
where the first error is statistical and the second systematic.
The sources of the systematic uncertainties are summarized in Table~\ref{src_sys_err}.
This measured branching fraction is consistent within error with
world average of $B(D^+ \to \mu^+\nu_{\mu})=(3.82 \pm 0.33)\times 10^{-4}$~\cite{pdg2010},
but with more precision.

    The decay constant $f_{D^+}$ can  be obtained
by inserting the measured branching fraction, the mass of the muon,
the mass of the $D^+$ meson, the CKM matrix element
$|V_{\rm cd}|= 0.2252\pm0.0007$ from the CKMFitter~\cite{pdg2010}
$G_F$
and the lifetime of the $D^+$ meson~\cite{pdg2010}
into Eq.(\ref{eq01}), which yields
$$f_{D^+} = (203.91 \pm 5.72 \pm 1.97)~~\rm MeV~~({\rm BESIII~Preliminary}),$$
\noindent
where the first errors are statistical and the second systematic arising
mainly from the uncertainties in
the measured branching fraction ($1.7\%$),
the CKM matrix element $|V_{\rm cd}|$ ($0.3 \%$),
and the lifetime of the $D^+$ meson ($0.7\%$)~\cite{pdg2010}.
The total systematic error is $1.0\%$.

\section{Leptonic decays of $D_s^+$ meson}

    The first observation of $D_s^+$ leptonic decay is performed 
at the WA75~\cite{WA75} fixed-target experiment in 1992. 
Since then experimental studies of the $D_s^+$ leptonic decays have been performed
at $e^+e^-$ experiments operated near $D_s^+D_s^-$ ($D_s^+D_s^{*-}$) production threshold,
at energies of the peak of $\Upsilon(2S)$ production and $Z^0$ production in $e^+e^-$
annihilation, as well as at other fixed-target experiments. In this section,
we review all of these available measurements of the $D_s^+$ leptonic decay branching fractions
and decay constants $f_{D_s^+}$.

\subsection{Results at fixed-target experiment}

\begin{figure}
\centerline{
\includegraphics[width=12.0cm,height=7.0cm]{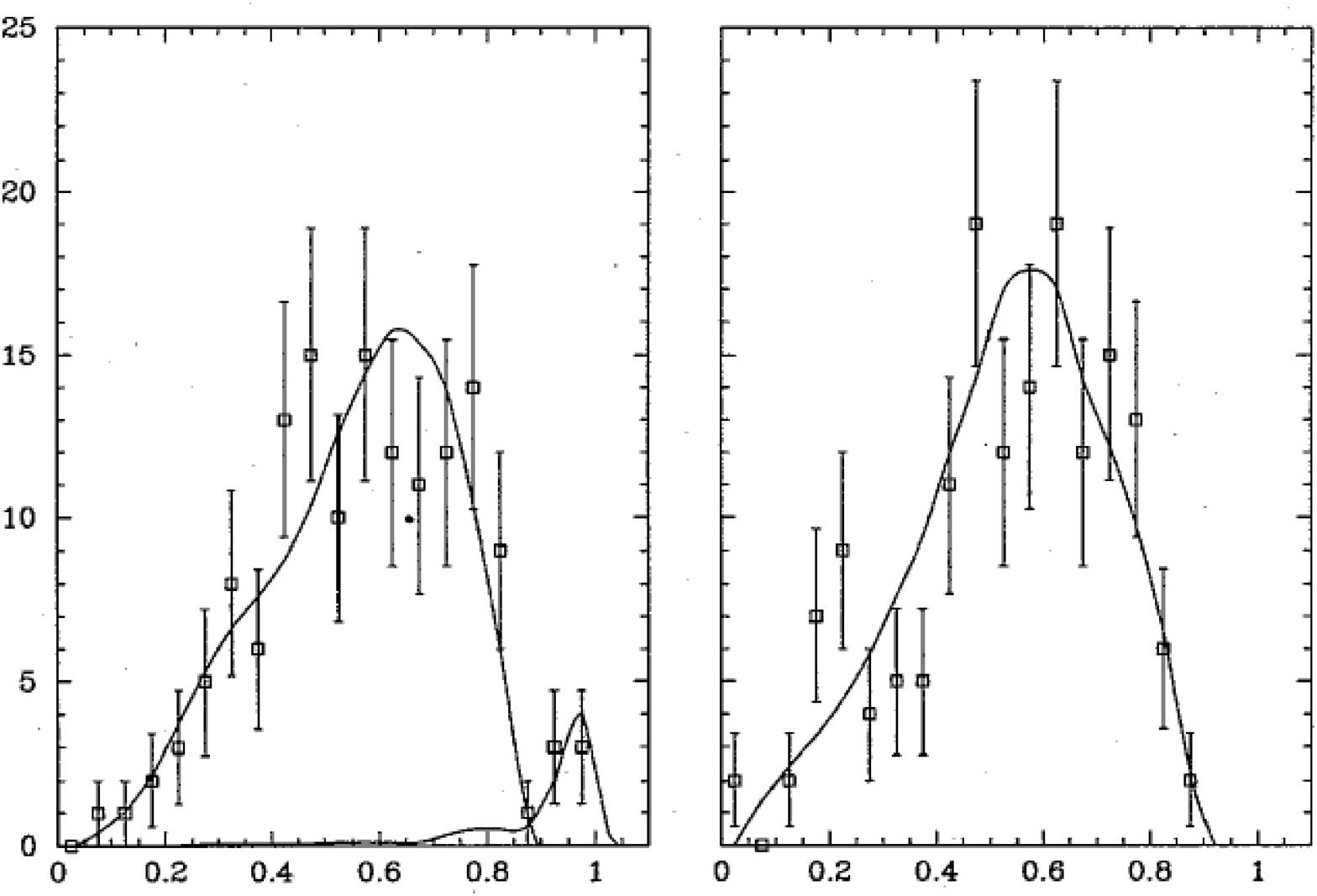}
\put(-190.0,-10.0){\large{$p_t^{\mu}$ GeV/$c$}}
\put(-200.0,165.0){(a)}
\put(-45.0,165.0){(b)}
        }
\caption{Distribution of muon momentum perpendicular to the direction of flight of the charm mesons
observed at the WA75~\cite{WA75} experiment, where (a) is for the candidates consistent with the decay of
the charm meson to a single charged particle, and (b) is for the candidates consistent with the decay of
the charm meson to two charged particles. The lines show the Monte Carlo predictions for these decays.}
\label{wa75_pmu_dstrs}
\end{figure}

   In 1992, the WA75~\cite{WA75} collaboration reported the first measurement of the branching fraction
for $D_s^+\rightarrow \mu^+\nu$ decay and measurement of the decay constant $f_{D_s^+}$.
To search for $D_s^+\rightarrow \mu^+\nu$ decay events 
they examined distribution of the muon momentum $p_t^{\mu}$ perpendicular
to the direction of flight of the charm mesons.
Figure~\ref{wa75_pmu_dstrs} (a) shows this momentum distribution for candidates consistent 
with the decay of a charged particle decaying to a single charged particle,
while Fig.~\ref{wa75_pmu_dstrs} (b) shows this momentum distribution for candidates consistent
with the decay of a neutral particle decaying to two charged particles.
The kinematic upper limit on $p_t^{\mu}$ is 0.98 GeV/$c$ for $D_s^+\rightarrow \mu^+\nu$
and 0.93 GeV/$c$ for $D^+\rightarrow \mu^+\nu$,
while the kinematic upper limit on $p_t^{\mu}$ is 0.88 GeV/$c$ for semileptonic decays.
With these different kinematic signatures of $p_t^{\mu}$ distributions at high transverse momentum
region, the leptonic decay of $D_s^+\rightarrow \mu^+\nu$ events can be well separated
from other background events of charm decays. By comparing the Fig.~\ref{wa75_pmu_dstrs} (a) 
and Fig.~\ref{wa75_pmu_dstrs} (b),
one can find that, in the charged topology, six events are observed with $p_t^{\mu}>0.9$ GeV/$c$,
while no event is observed above $p_t^{\mu}>0.9$ GeV/$c$ in the neutral topology.
The estimated number of the background events from $D^+\rightarrow \mu^+\nu$ decay
is $0.6\pm 0.2$ events. Based on these six candidate events for  $D_s^+\rightarrow \mu^+\nu$ decay,
$0.6\pm 0.2$  background events from $D^+\rightarrow \mu^+\nu$ and the number of events of
$D^0 \rightarrow \mu^+\nu X$ for normalization, the WA75~\cite{WA75} collaboration determined a branching fraction
of $B(D_s^+\rightarrow \mu^+\nu)=(4.0^{+1.8+0.8}_{-1.4-0.6}\pm1.7)\times 10^{-3}$,
and a decay constant of $f_{D_s^+}=(225 \pm 45 \pm 20 \pm 40)$ MeV~\cite{WA75}.

    As WA75 experiment, the Fermilab E635~\cite{E653} is a fixed-target experiment with an emulsion target and muon trigger.
In 1996, the E653~\cite{E653} collaboration observed 23 events for $D^+_s\rightarrow \mu^+\nu$ leptonic decays
in the fixed-target experiment. Based on the yields of $D_s^+\rightarrow \phi\mu^+\nu$ signal 
observed in the same data sample, the E653~\cite{E653} collaboration determined a relative decay branching fraction and
decay constant of $B(D_s^+\rightarrow \mu^+\nu)/B(D_s^+\rightarrow \phi\mu^+\nu)=(0.16 \pm 0.06 \pm 0.03)$ and
$f_{D_s^+}=(194 \pm 35 \pm 20 \pm 14)$ MeV~\cite{E653}, respectively.

    In 2000, using almost the same analysis technique as the one used by WA75~\cite{WA75},
the BEATRICE collaboration observed $D_s^+\rightarrow \mu^+\nu$ leptonic decays at the WA92 experiment. 
They measured a relative decay branching fraction and decay constant of 
$B(D_s^+\to\mu^+\nu)/B(D_s^+\to\phi_{(K^+K^-)}\pi^+)=0.47\pm0.13\pm0.04\pm0.06$ and 
$f_{D_s^+}=(323 \pm 44 \pm 12 \pm 34)$ MeV~\cite{WA92}, respectively.

\subsection{Results at $e^+e^-$ experiments near $D_s^+D_s^-$ thresholds}

\subsubsection{BES-I experiment near $D_s^+D_s^-$ threshold}

   In 1995, by analyzing the data taken  at 4.03 GeV with the BES-I detector operated at the BEPC collider,
the BES collaboration reconstructed $94.3\pm 12.5$ singly tagged $D_s^-$ mesons with three hadronic decay modes.
In the system recoiling against the singly tagged $D_s^-$ mesons, the BES collaboration found 3 events of both
the $D_s^+\rightarrow \tau^+\nu$ and $D_s^+\rightarrow \mu^+\nu$. They measured the decay branching fractions
for $D_s^+\rightarrow \tau^+\nu$ and $D_s^+\rightarrow \mu^+\nu$ to be 
$B(D_s^+\rightarrow \tau^+\nu)=(15^{+13+3}_{-6-2})\%$ and
$B(D_s^+\rightarrow \mu^+\nu)=(1.5^{+1.3+0.3}_{-0.6-0.2})\%$, respectively. They
extracted a value of the decay constant
$f_{D_s^+}=(430^{+150}_{-130} \pm 40)$ MeV~\cite{bes-i_fDs}, 
where the first error is statistical and the second is the systematic uncertainty arising from the uncertainties
of reconstruction efficiency, background estimation and the $D_s^+$ lifetime. 

These are the first absolute measurements of these decay branching fractions and decay constant, 
which do not need to normalize to other $D_s^+$ decay modes and
do not depend on the knowing $D_s^+$ production rate in the data samples.

\subsubsection{CLEO-c experiment near $D_s^+D_s^{*-}$ threshold}

In 2009, the CLEO-c analyzed 600 pb$^{-1}$ of data taken at 4.17 GeV in $e^+e^-$ annihilation
to measure branching fractions for $D_s^+\rightarrow l^+\nu$ decays
and decay constant $f_{D_s^+}$. 
From this data sample, they accumulated the singly tagged $D_s^-$ mesons
using 9 hadronic decay modes. 
Since the $D_s^+D_s^{*-}$meson pairs are produced in $e^+e^-$ collision, 
the CLEO collaboration used the missing mass square method to reconstruct the decay 
of $D_s^{*-} \rightarrow \gamma D_s^-$. 
They calculate the variable $MM^{*2} = (E_{\rm CM}-E_{D_s^-}-E_{\gamma})^2-(p_{\rm CM}-p_{D_s^-}-p_{\gamma})^2$
for each event. $MM^{*2}$ is the missing mass-squared of the system recoiling against the $D_s^{*-}$.
With the peak of $MM^{*2}$ distributions for each of nine decay modes, they clearly  reconstructed the $D_s^+$ mesons.
By fitting to these $MM^{*2}$ distributions, they obtained the number of the $D_s^+$ mesons in total.
In the system recoiling against the singly tagged
$D_s^{*-}$, they searched for the leptonic decays of $D_s^+ \rightarrow \mu^+\nu$ and
$D_s^+ \rightarrow \tau^+\nu$. Finally, they measured the decay branching fractions of
$B(D_s^+ \rightarrow \mu^+\nu)=(0.565 \pm 0.045 \pm 0.017)\%$ and
$B(D_s^+ \rightarrow \tau^+\nu)=(6.42 \pm 0.81 \pm 0.18)\%$,
and extracted the decay constant of $f_{D_s^+}=(263.3 \pm 8.2 \pm 3.9)$ MeV~\cite{cleo-c_fDs_2009a}.

Using three cleanest singly tagged $D_s^-$ hadronic decay modes to accumulate the $D_s^-$ tags, 
the CLEO collaboration searched for $D_s^+ \rightarrow \tau^+ \nu \rightarrow e^+\nu\nu\nu$ decays.  
They measured the decay branching fraction and decay constant of 
$B(D_s^+\to\tau^+\nu)=(5.30\pm0.47\pm0.22)\%$ and 
$f_{D_s^+}=(252.5 \pm 11.1 \pm 5.2)$ MeV~\cite{cleo-c_fDs_2009b}, respectively.

In addition to the $D_s^+ \rightarrow \tau^+ \nu \rightarrow e^+\nu\nu\nu$ decay mode,
the CLEO also observed $D_s^+ \rightarrow \tau^+ \nu \rightarrow \rho^+\nu\nu$ decays.
With this decay process the CLEO measured the $D^+\rightarrow \tau^+\nu$ decay branching fraction,
which is  $B(D_s^+ \rightarrow \tau^+\nu)=(5.52 \pm 0.57 \pm 0.21)\%$. With this decay branching fraction,
they extracted the decay constant of $f_{D_s^+}=(257.8 \pm 13.3 \pm 5.2)$ MeV~\cite{cleo-c_fDs_2009c}. 
  
\subsection{Results at $e^+e^-$ experiments operated at higher energies}

As we mentioned before, the $D_s^+$ meson can also be produced from the quark fragmentation process
in continuum $c\bar c$ production as well as produced in $Z^0$ decays. So the $D_s^+$ leptonic decays
can be studied with the large data samples taken at $\Upsilon(4S)$ production energy 
and taken at $Z^0$ production energy.
In this section we review the results
on measurements of branching fractions of $D_s^+$ leptonic decays and decay constant $f_{D_s^+}$
measured at these two energies.

\subsubsection{Results at $e^+e^-$ experiments operated at 10.5 GeV}

Several $e^+e^-$ experiments operated at 10.5 GeV have studied or have been studying the $D_s^+$ leptonic decays.

In 1998, the CLEO-II observed $182 \pm 22$ events for $D_s^{*+} \rightarrow \gamma D_s^+$
followed by $D_s^{+} \rightarrow \mu^+ \nu$ by analyzing 
using 5 million $e^+e^-\to c\bar c$ events. They measured the decay width ratio of 
$\Gamma(D_s^+\to\mu^+\nu)/\Gamma(D^+_s\to\phi\pi^+) = 0.173\pm0.023\pm0.035$ 
and determined the decay constant 
of $f_{D_s^+}=(280 \pm 19 \pm 28 \pm 34)$ MeV~\cite{cleo_y1994_fDs}. 

In 2008, BELLE collaboration made measurements of leptonic decay branching fractions
of $D_s^+$ meson. 
They selected the $D_s^+$ leptonic decays from the $e^+e^- \rightarrow c\bar c$ continuum production, 
during which the $D_s^{*}D^{\pm,0}K^{\pm,0}X$ produced from the quark fragmentation,
where $D_s^* \rightarrow \gamma D_s$ and $X$ indicates several pions or photons.
By reconstructing the recoil mass of the $DKX \gamma$, they observed clear
$D_s$ signal in the system recoiling against the $DKX \gamma$. By fitting the mass distributions
of the system recoiling against $DKX \gamma$, they accumulated $32100\pm 870 \pm 1210$ $D_s$ events.
Then they examined the mass distribution of the system recoiling against the
$DKX \gamma \mu$ combinations. They found a very clear signal for 
$D_s^{+} \rightarrow l^+ \nu$ decays with a narrow peak around 0.0 in the
missing mass squared $M^2_{\rm REC}(DKX \gamma \mu)$ distribution. 
Fitting this $M^2_{\rm REC}(DKX \gamma \mu)$ distribution yields $169 \pm 16 \pm 8$
signal events for $D_s^{+} \rightarrow l^+ \nu$ decays. With these numbers, 
the BELLE collaboration measured the decay branching fraction
of $B(D_s^+ \rightarrow \mu^+\nu)=(0.644 \pm 0.076 \pm 0.057)\%$,
and decay constant of $f_{D_s^+}=(275 \pm 16 \pm 12)$ MeV~\cite{belle_y2008_fDs}.

At the Charm2012 Conference, the BELLE collaboration presented an updated analysis 
of their 913 fb$^{-1}$ of data collected at 10.6 GeV.
With a larger data sample, the BELLE collaboration observed $489 \pm 26$ signal events for 
$D_s^+ \rightarrow \mu^+\nu$ decay and
measured the decay branching fraction
of $B(D_s^+ \rightarrow \mu^+\nu)=(0.528 \pm 0.028 \pm 0.019)\%$.
In addition to this decay, the BELLE collaboration observed 
$2206 \pm 84$ 
signal events for $D_s^+ \rightarrow \tau^+\nu$ with the decays of $\tau^+\rightarrow e^+\nu\nu$,
$\tau^+\rightarrow \mu^+\nu\nu$ and  $\tau^+\rightarrow \pi^+\nu$, and they measured
the decay branching fraction of $B(D_s^+ \rightarrow \tau^+\nu)=(5.70 \pm 0.21^{+0.31}_{-0.30})\%$. 
With these two decay modes together,
they extracted the decay constant 
of $f_{D_s^+}=(255.0 \pm 4.2 \pm 5.0)$ MeV~\cite{belle_charm2012_fDs}.

In 2010, using the same technique as the one used by the BELLE collaboration,
the BaBar collaboration made measurements of the $D_s$ leptonic decay branching fractions
and determined decay constant.
By analyzing 521 fb$^{-1}$ of data taken at 10.6 GeV,
the BaBar collaboration measured the decay branching fractions 
for $D_s^+ \rightarrow \mu^+\nu$, $D_s^+ \rightarrow \tau^+\nu$ ($\tau^+ \rightarrow e^+\nu\nu$),
and $D_s^+ \rightarrow \tau^+\nu$ ($\tau^+ \rightarrow \mu^+\nu\nu$) to be
$B(D_s^+ \rightarrow \mu^+\nu)=(0.602 \pm 0.038 \pm 0.034)\%$,
$B(D_s^+ \rightarrow \tau^+\nu)=(5.07 \pm 0.52 \pm 0.68)\%$ and
$B(D_s^+ \rightarrow \tau^+\nu)=(4.91 \pm 0.47 \pm 0.54)\%$, respectively, 
and determined the decay constant of
$f_{D_s^+}=(258.6 \pm 6.4 \pm 7.5)$ MeV~\cite{babar_y2008_fDs}.

\subsubsection{Results at $e^+e^-$ experiments operated at 91 GeV}

In $e^+e^-$ annihilation at 91 GeV, the $Z^0$ bosons are produced. The $Z^0$ boson can decay into $c\bar c$.
Due to quark fragmentation the $D_s^+$ meson are produced in the final states of
the $Z^0$ decays.
At these experiments, the decays of $D_s^+ \rightarrow l^+ \nu$ are selected by
reconstructing the decay sequence of
$e^+e^- \rightarrow Z^0 \rightarrow c\bar c \rightarrow D_s^{*-}X$, 
where $D_s^{*-} \rightarrow \gamma D_s^-$
with $D_s^- \rightarrow l^-\nu$.

In 1997, L3 collaboration observed $15.5\pm 6.0$ $D_s^+ \rightarrow \tau^+\nu$ events 
coming from $1.5 \times 10^6$ $Z^0 \rightarrow q\bar q(\gamma)$ events.
They measured the leptonic decay branching fraction
of $B(D_s^+ \rightarrow \tau^+\nu)=(7.4 \pm 2.8 \pm 1.6 \pm 1.8)\%$, and
decay constant of $f_{D_s^+}=(309 \pm 58 \pm 33 \pm 38)$ MeV~\cite{l3_fDs}.

In 2001, the OPAL collaboration observed 
$22.5\pm 6.9$ $D_s^+ \rightarrow \tau^+\nu$ events
coming from $3.9 \times 10^6$ $Z^0 \rightarrow q\bar q(\gamma)$ events.
They measured the leptonic decay branching fraction
of $B(D_s^+ \rightarrow \tau^+\nu)=(7.0 \pm 2.1 \pm 2.0)\%$, and
decay constant of $f_{D_s^+}=(286 \pm 44 \pm 41)$ MeV~\cite{opal_fDs}. 

In 2002, the ALEPH collaboration made a measurement of leptonic decay branching fractions
and decay constant of $D_s^+$ meson. The measurements were made based on an almost
same technique as the one used by the L3 and OPAL collaborations. But the ALEPH reconstructed
the $D_s^+ \rightarrow \mu^+\nu$ decay directly. 
By analyzing $3.97 \times 10^6$ hadronic $Z^0$ decays, 
they measured the leptonic decay branching fractions
of $B(D_s^+ \rightarrow \tau^+\nu)=(5.79 \pm 0.77 \pm 1.84)\%$ and
$B(D_s^+ \rightarrow \mu^+\nu)=(0.68 \pm 0.11 \pm 0.18)\%$,
and decay constant of $f_{D_s^+}=(285 \pm 19 \pm 40)$ MeV~\cite{aleph_fDs}.

\section{Comparison of measured and expected decay constants of $f_{D^+}$ and $f_{D_{s}^+}$}

\subsection{Re-determine $f_{D_{(s)}^+}$}
 
    The values of the decay constants of $f_{D^+}$ and $f_{D_s^+}$ 
measured at different experiments
were historically obtained with the measured leptonic decay branching fractions at these experiments,
with the lifetimes and masses of the $D^+$ and $D_s^+$ mesons,
together with the CKM matrix elements of $|V_{\rm cd}|$ and $|V_{\rm cs}|$, 
or with the measured branching fractions for $D_s^+ \rightarrow \phi \pi^+$ decay
or other decays as inputs. 
The historical values of the lifetimes, the CKM matrix elements 
and branching fractions for $D_s^+ \rightarrow \phi \pi^+$ decay
or other decays
used in determination of the values of the $f_{D^+}$ and $f_{D_s^+}$
differ from each at these experiments. In order to
make precise comparison of these measured decay constants, 
we re-calculate the decay constants based on the originally measured branching fractions
for these two leptonic decays of the $D^+$ and $D_s^+$ mesons.
    In re-determination of the decay constants $f_{D^+}$ and $f_{D_s^+}$,   
the values of physical quantities used are listed in Table~\ref{physics_quantities},
which are quoted from PDG2010~\cite{pdg2010}.
 \begin{table}[hbp]
  \caption{The values of physical quantities used in the re-determination of
   $f_{D^+}$ and $f_{D^+_s}$.}
  \label{physics_quantities}
  \resizebox{1.0\textwidth}{!}{
  \begin{tabular}{|c|c|c|c|}
   \hline
    $D^+_{(s)}$ mass & $D^+_{(s)}$ lifetime & lepton mass & $|V_{cd}|$ or $|V_{cs}|$     \\
   \hline
    $m_{D^+} = (1869.60\pm0.16)~\rm MeV$ & $\tau_{D^+} = (1040\pm7)\times 10^{-15}~\rm s$ & 
                                           $m_{\mu} = (105.658367\pm0.000004)~\rm MeV$ & $|V_{cd}| = 0.2252\pm0.0007$ \\
   \hline
    $m_{D^+_s} = (1968.47\pm0.33)~\rm MeV$ & $\tau_{D^+_s} = (500\pm7)\times 10^{-15}~\rm s$ & 
                                             $m_{\tau} = (1776.82\pm0.16)~\rm {MeV}$ & $|V_{cs}| = 0.97345^{+0.00015}_{-0.00016}$ \\
   \hline
  \end{tabular}}
 \end{table}

\subsection{Comparison of the measured and expected $f_{D_{(s)}^+}$}

    Decay constants for pseudoscalar mesons containing a heavy $c$ and/or $b$ quark have been predicted
with theories or models based on the QCD. 
In recent years, the LQCD calculations of the decay constants $f_{D_{(s)}^+}$
have achieved high precision. 
Some theoretical predictions for the decay constants were calculated 
in Refs.~\cite{lqcd_HPQCD_UKQCD,Amundson_prd47_p3059_y1993}.

\begin{figure}
\centerline{
\includegraphics[width=9.0cm,height=9.0cm]{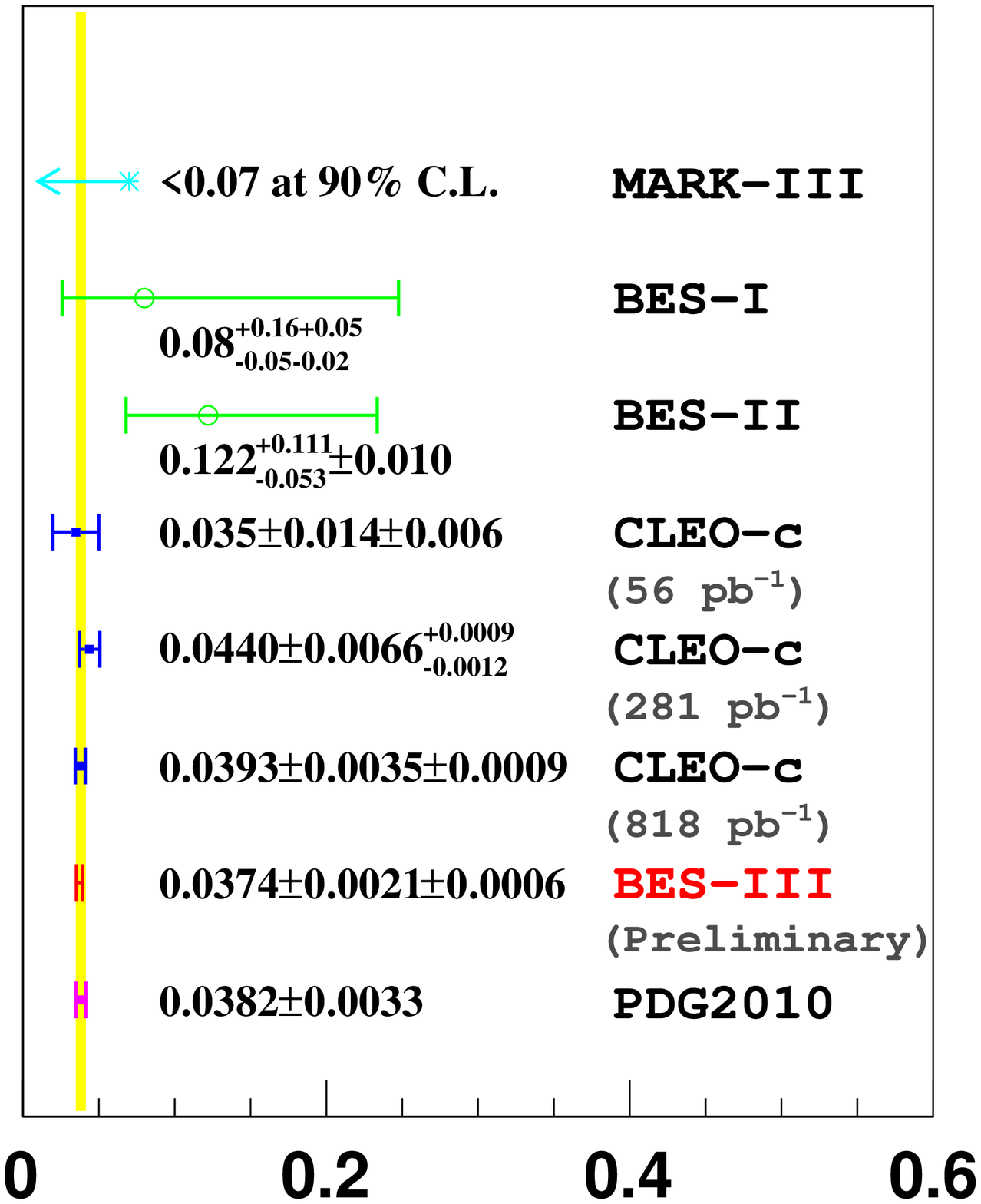}
\put(-180.0,12.0){\large{$B(D^+ \rightarrow \mu^+\nu)~[\%]$}}
\put(-50.0,220.0){(a)}
\includegraphics[width=9.0cm,height=9.0cm]{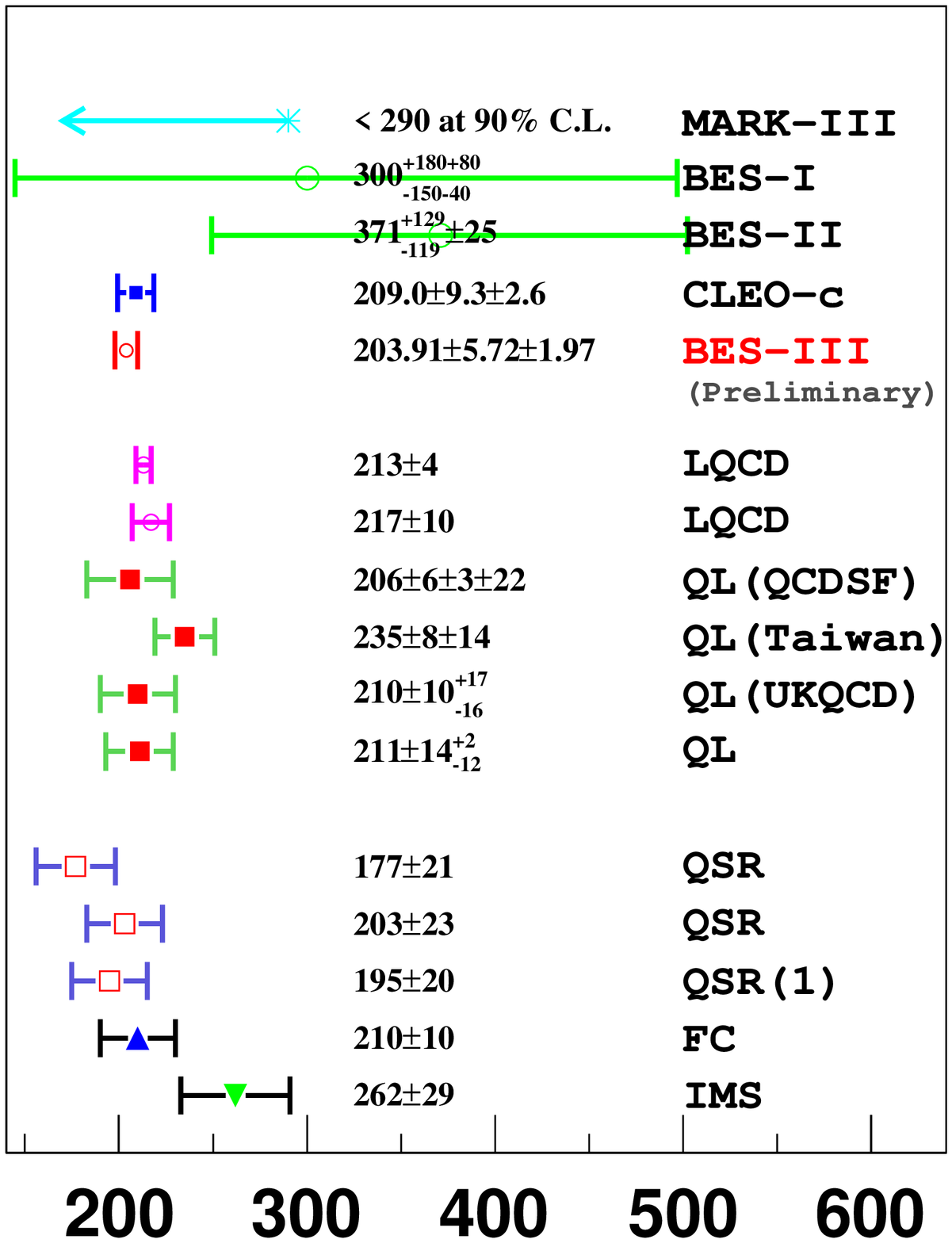}
\put(-160.0,12.0){\large{$f_{D^+}$ [MeV]}}
\put(-50.0,220.0){(b)}
        }
\caption{(a) Comparison of the measured branching fractions for $D^+ \rightarrow \mu^+\nu$ decay,
and (b) comparison of the measured decay constant $f_{D^+}$ and 
these predicted with theories~\cite{lqcd_HPQCD_UKQCD,Amundson_prd47_p3059_y1993},
where QSR is for QCD Sum Rule, FC is for Field Correlations, and IMS is for Isospin Mass Splittings.} 
\label{fig4}
\end{figure}

Figure~\ref{fig4} (a) and (b) give comparison of the measured branching fractions for $D^+ \rightarrow \mu^+\nu$
and comparison of the measured values of the decay constant $f_{D^+}$
with those predicted with different theoretical calculations, respectively.
The weighted average of the predicted values of decay constant $f_{D^+}$ with theories based on QCD
is $f_{D^+}=(212.7\pm 3.2)$ 
MeV~\footnote{We did not use the predicted values given by QSR(1) and IMS in calculating
the weighted average of the predicted values
of decay constant $f_{D^+}$ since the ratios of $f_{D_s^+}/f_{D^+}$ are not available in Refs.}.
\begin{figure}
\centerline{
\includegraphics[width=9.0cm,height=9.0cm]{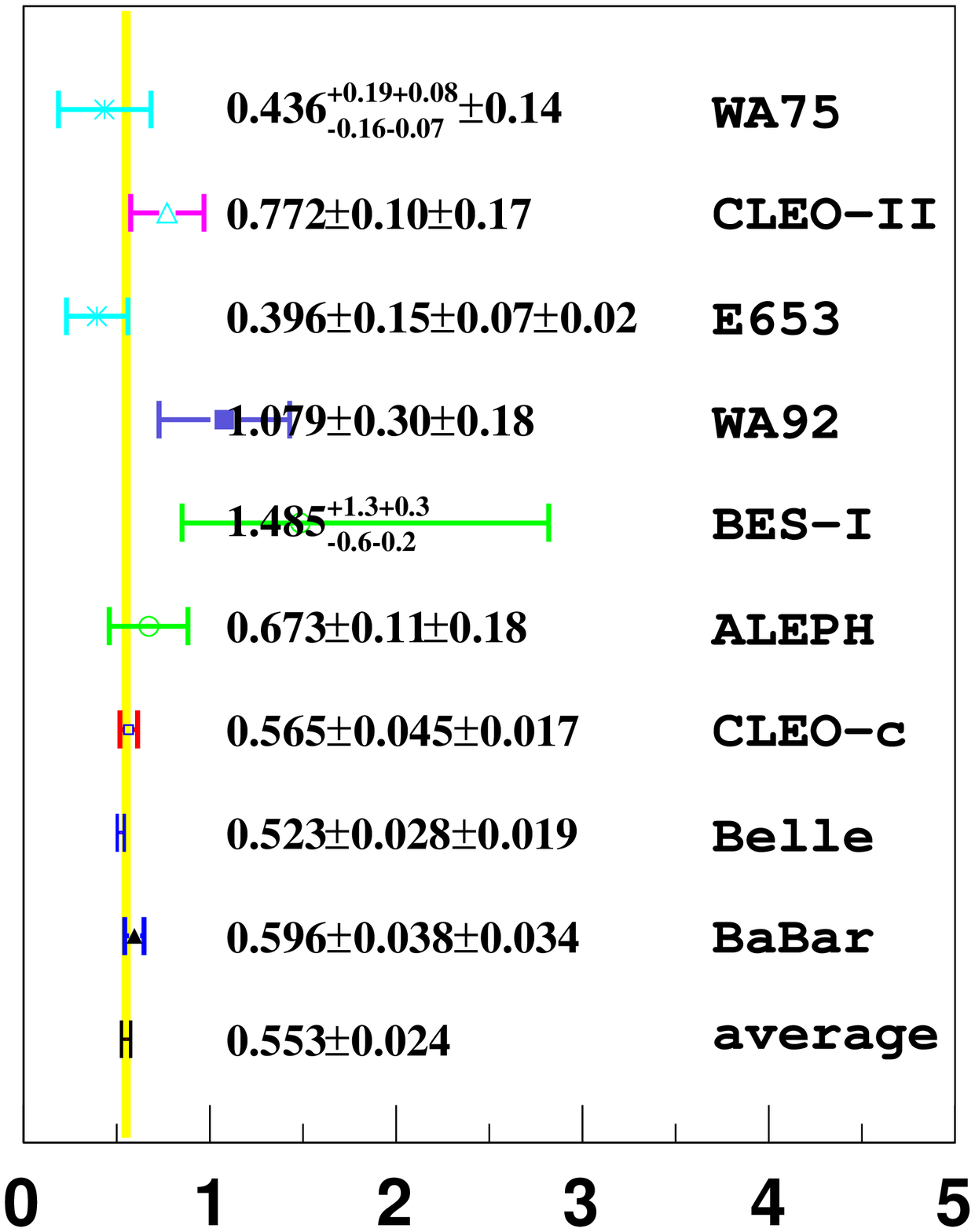}
\put(-180.0,12.0){\large{$B(D_s^+ \rightarrow \mu^+\nu)~[\%]$}}
\put(-50.0,220.0){(a)}
\includegraphics[width=9.0cm,height=9.0cm]{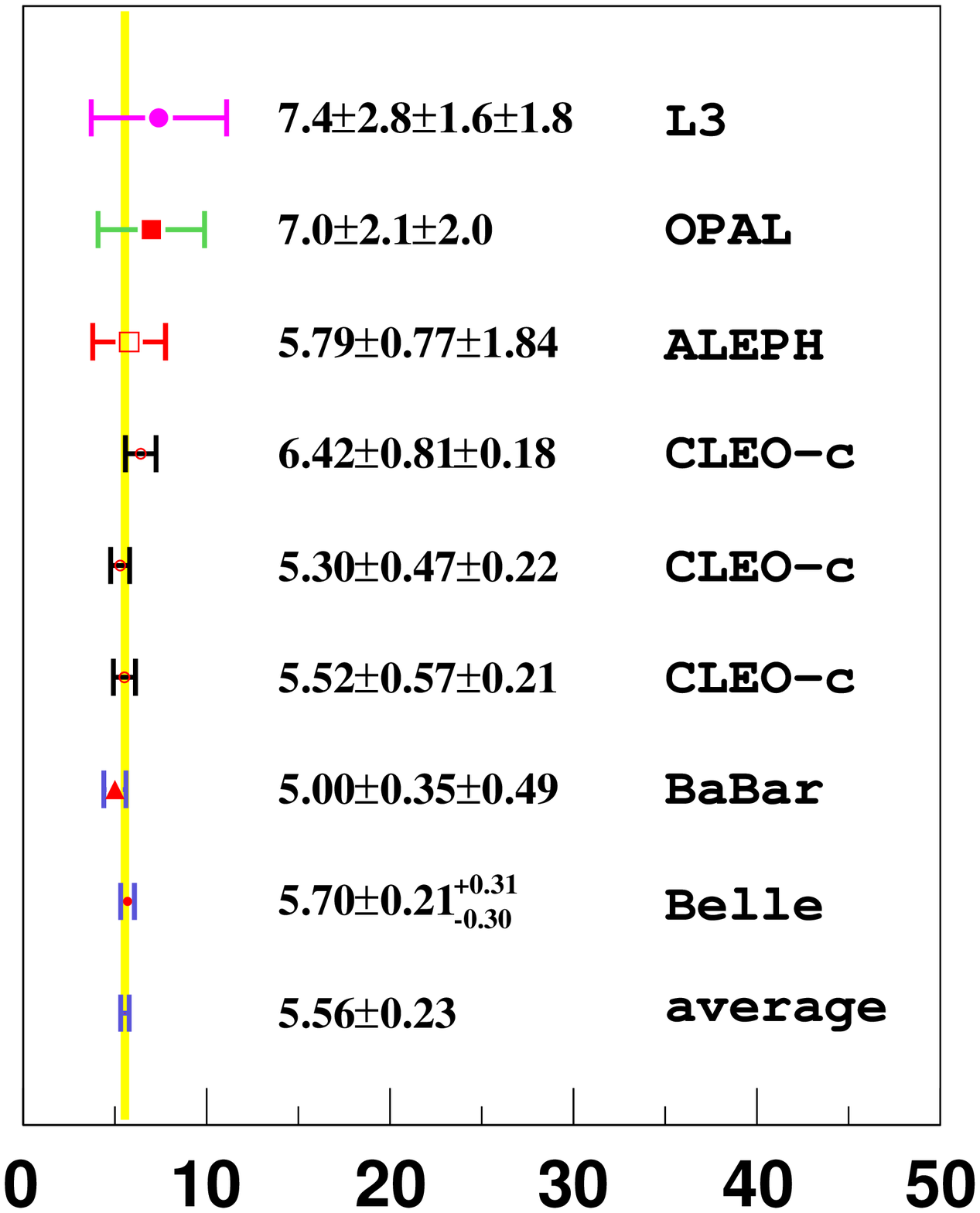}
\put(-170.0,12.0){\large{$B(D_s^+ \rightarrow \tau^+\nu)~[\%]$}}
\put(-50.0,220.0){(b)}
        }
\caption{Comparison of the measured branching fractions for (a) $D_s^+ \rightarrow \mu^+\nu$ decay and
(b) $D_s^+ \rightarrow \tau^+\nu$ decay.}
\label{fig5}
\end{figure}

   Figure~\ref{fig5} (a) and (b) show comparison of the branching fractions for $D_s^+ \rightarrow \mu^+\nu$
and $D_s^+ \rightarrow \tau^+\nu$ decays measured at different experiments, respectively. 
With these measured decay branching fractions, 
we obtain the average branching fraction for $D_s^+\rightarrow \mu^+\nu$
and $D_s^+\rightarrow \tau^+\nu$ decays to be $B(D_s^+\rightarrow \mu^+\nu)=(0.553\pm 0.024)\%$ and
$B(D_s^+\rightarrow \tau^+\nu)=(5.56\pm 0.23)\%$, respectively.

\begin{figure}
\centerline{
\includegraphics[width=9.0cm,height=9.0cm] {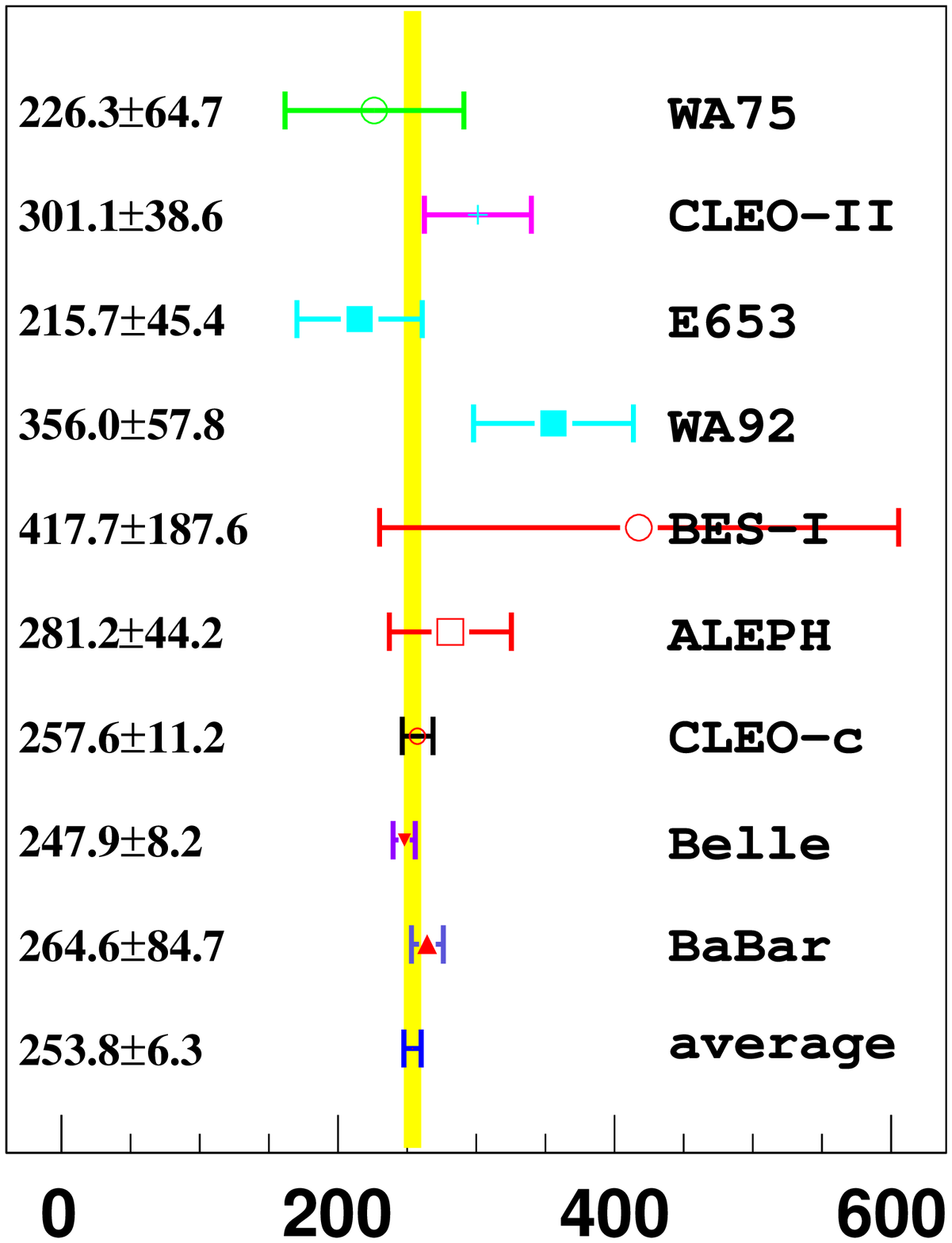}
\put(-170.0,12.0){\large{$f_{D_s^+}$ [MeV] }}
\put(-50.0,220.0){(a)}
\put(-160.0,235.0){$D_s^+\rightarrow \mu^+\nu$}
\includegraphics[width=9.0cm,height=9.0cm] {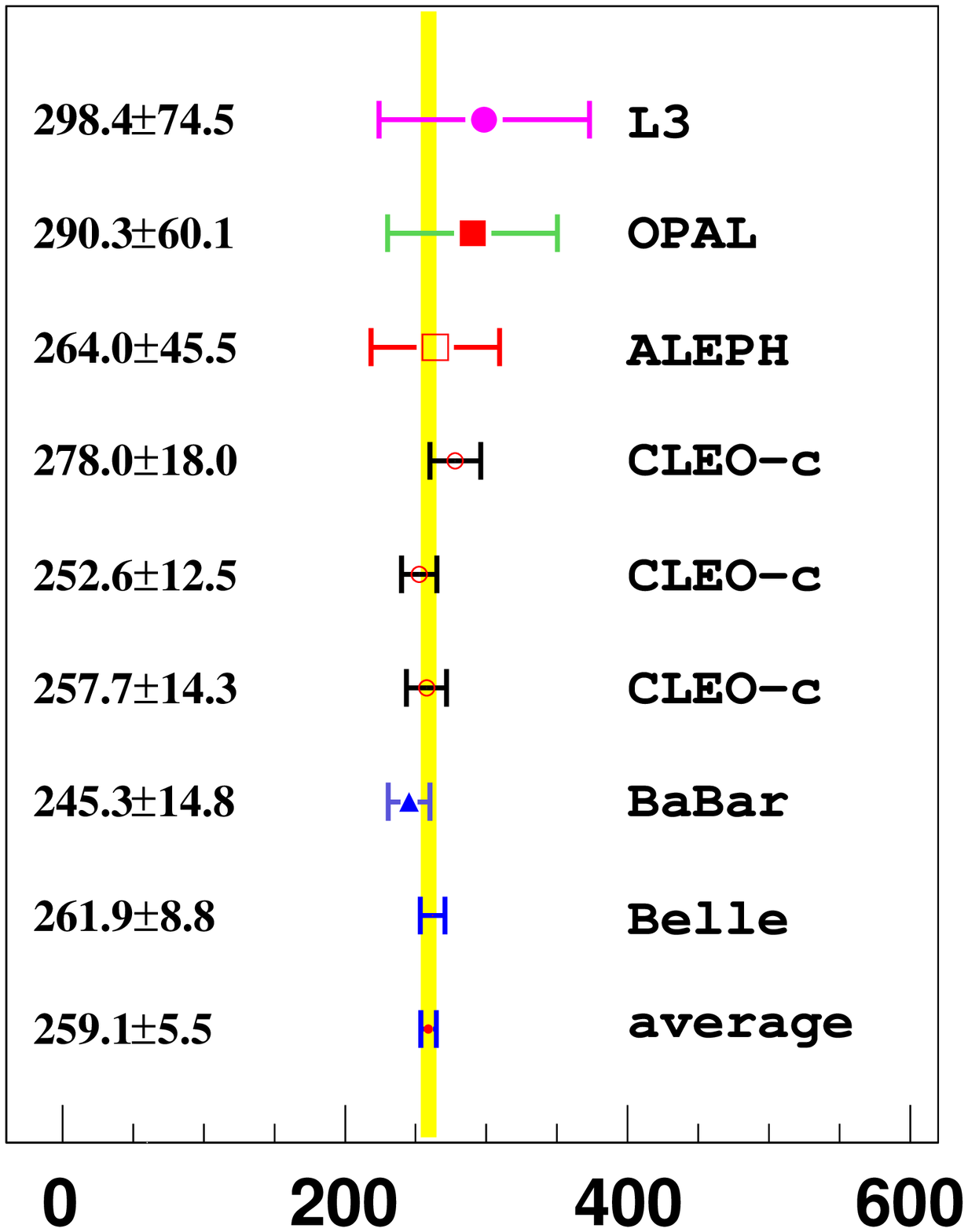}
\put(-160.0,12.0){\large{$f_{D_s^+}$ [MeV]}}
\put(-50.0,220.0){(b)}
\put(-160.0,235.0){$D_s^+\rightarrow \tau^+\nu$}
        }
\caption{Comparison of the measured $f_{D_s^+}$ obtained from (a) $D_s^+ \rightarrow \mu^+\nu$ decay
and (b) $D_s^+ \rightarrow \tau^+\nu$ decay.}
\label{fig6}
\end{figure}

\begin{figure}
\centerline{
\includegraphics[width=9.0cm,height=9.0cm]
{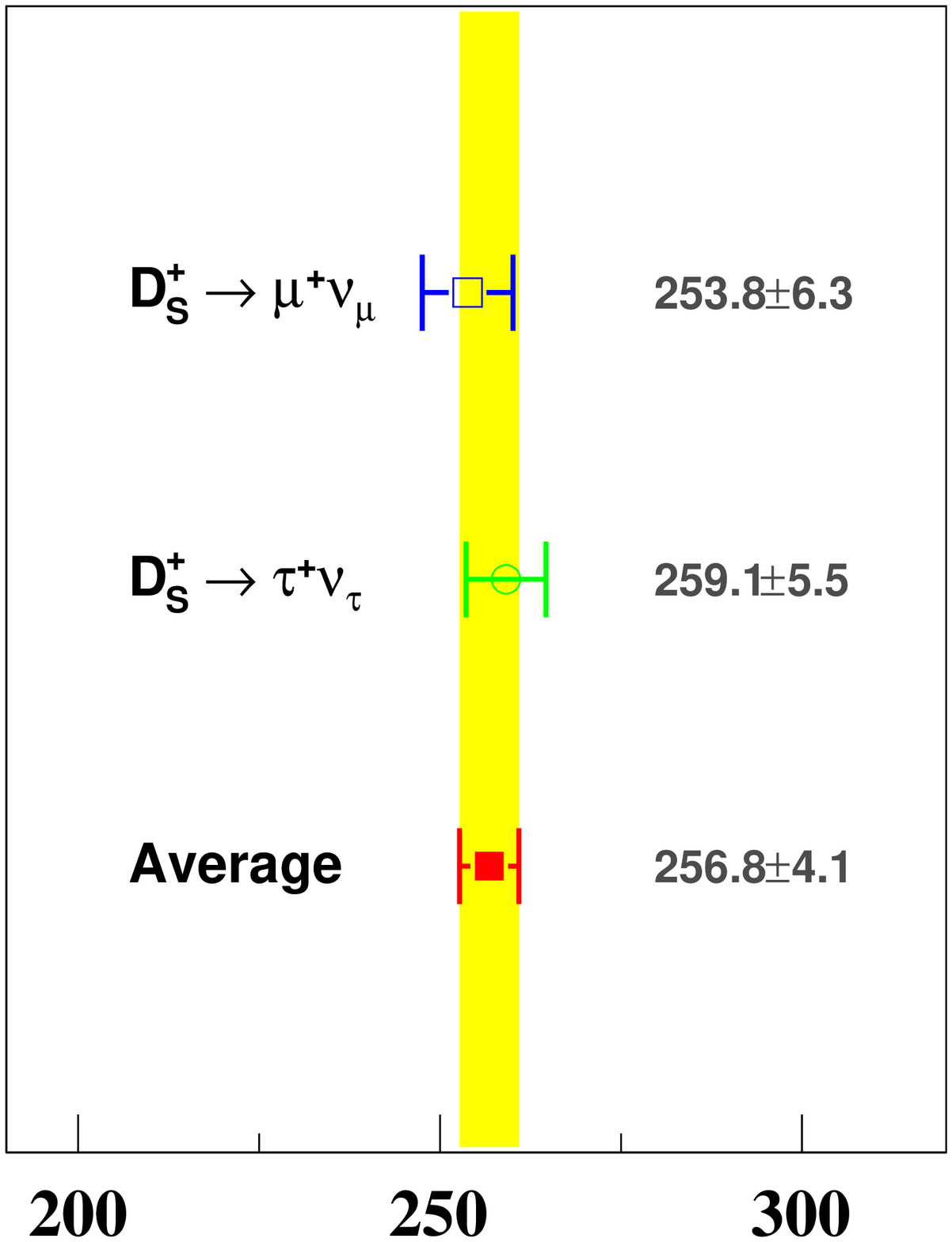}
\put(-170.0,12.0){\large{$f_{D_s^+}$ [MeV] }}
\put(-50.0,220.0){(a)}
\includegraphics[width=9.0cm,height=9.0cm]
{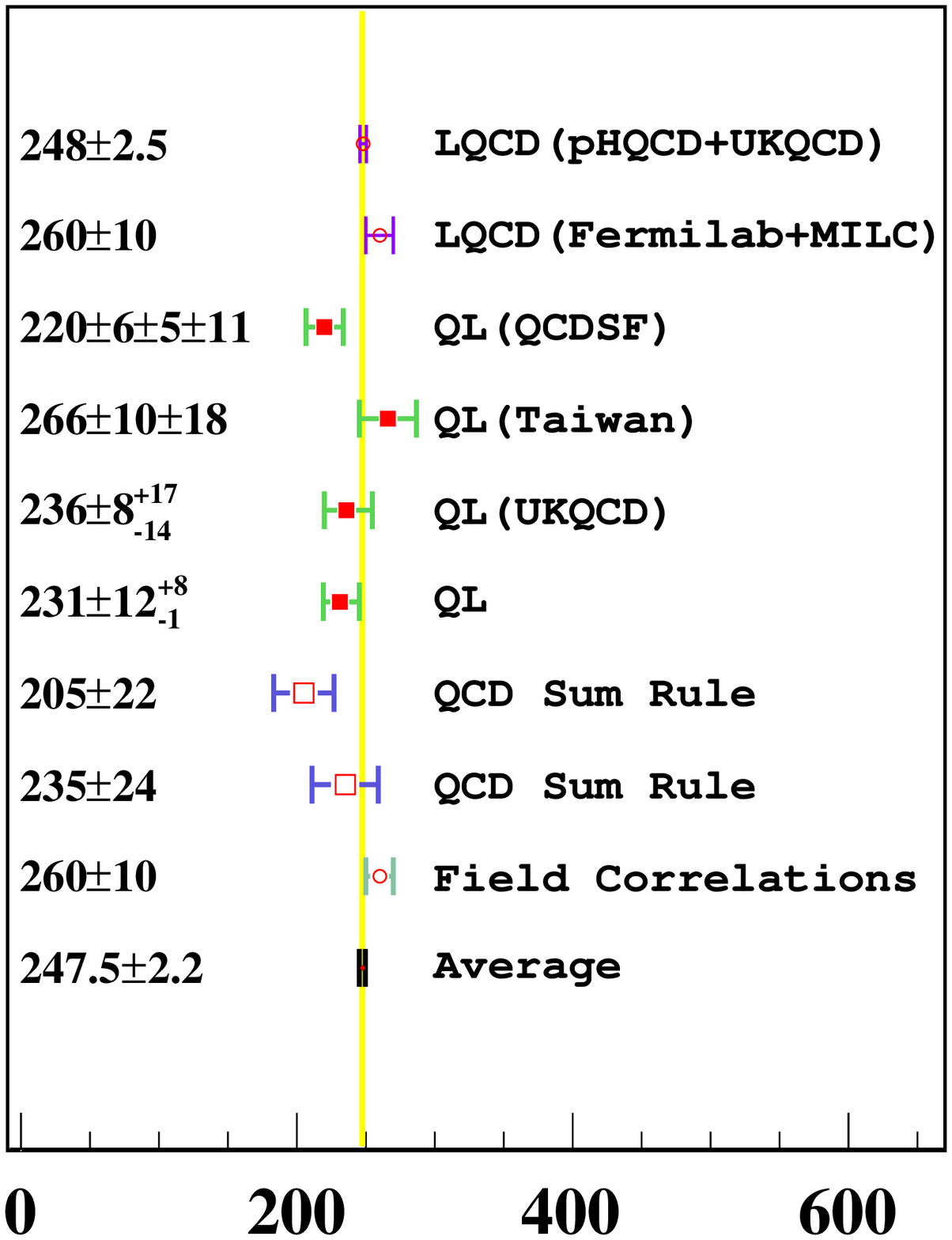}
\put(-160.0,12.0){\large{$f_{D_s^+}$ [MeV]}}
\put(-50.0,220.0){(b)}
        }
\caption{Comparison of (a) the measured $f_{D_s^+}$ obtained from $D_s^+ \rightarrow \mu^+\nu$ decay
and $D_s^+\rightarrow \tau^+\nu$ decay, (b) the predicted $f_{D^+_s}$ with theories based on QCD.}
\label{fig7}
\end{figure}

   With these branching fractions for the two leptonic decays measured at the different experiments, 
we obtain the values of the decay constant $f_{D_s^+}$.    
Figure~\ref{fig6} (a) and (b) show comparison of the measured values of $f_{D_s^+}$
obtained from the decays of $D_s^+ \rightarrow \mu^+\nu$ and $D_s^+ \rightarrow \tau^+\nu$, respectively.
Weighting these measured values of $f_{D_s^+}$ with these errors yields the decay constant
of $f_{D_s^+}=(253.8 \pm 6.3)$ MeV and $f_{D_s^+}=(259.1 \pm 5.5)$ MeV obtained from
the decays of $D_s^+\rightarrow \mu^+\nu$ and $D_s^+\rightarrow \tau^+\nu$, respectively.
Figure~\ref{fig7} (a) shows a comparison of these two averaged 
decay constant $f_{D_s^+}$ obtained from $D_s^+ \rightarrow \mu^+\nu$
and $D_s^+ \rightarrow \tau^+\nu$ decays and the weighted average of the two decay constants. 
Figure~\ref{fig7} (b) shows comparison of values of $f_{D_s^+}$ predicted with theories.
Weighted the values of $f_{D_s^+}$ predicted with theories based on QCD with their errors
yields an average value of the theoretically predicted decay constants, 
which is $f_{D_s^+}=(247.5 \pm 2.2)$ MeV.

\subsection{Comparison of measured and expected ratio of $f_{D_s^+}/f_{D^+}$}

\begin{figure}
\centerline{
\includegraphics[width=9.0cm,height=9.0cm]
{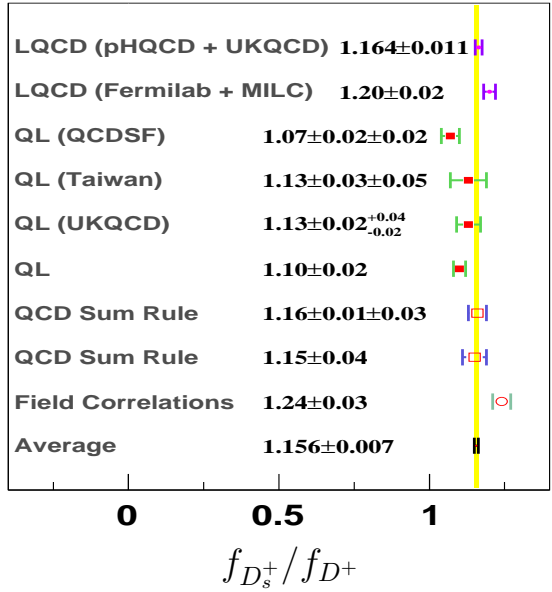}
\put(-150.0,14.0){\large{$f_{D_s^+}/f_{D^+}$ }}
        }
\caption{Comparison of the ratio of $f_{D_s^+}/f_{D^+}$ predicted with theories based on QCD.}
\label{fig8}
\end{figure}

    By weighting the decay constant $f_{D^+}$ measured at the CLEO-c and BES-III experiments,
we obtain $f_{D^+}=205.3 \pm 5.1$ MeV. While the weighted average of the values of the measured
$D_s^+$ decay constant is $f_{D_s^+}=256.8 \pm 4.1$ MeV. 
With those two measured decay constants, we obtain the ratio
$f_{D_s^+}/f_{D^+}=1.251 \pm 0.037$.
Figure~\ref{fig8} shows comparison of the predicted ratio $f_{D_s^+}/f_{D^+}$ 
with different theories based on QCD. The weighted average of these ratios is $f_{D_s^+}/f_{D^+}=1.156 \pm 0.007$.
At present, the measured ratio of $f_{D_s^+}/f_{D^+}$ is 2.5$\sigma$ larger than the one predicted 
with theoretical calculations. 
This 2.5$\sigma$ deviation of the measured ratio of $f_{D_s^+}/f_{D^+}$ from the predicted ratio 
with theories based on QCD may indicate that some effects of non-standard model enhance the decay rate of
$D_s^+ \rightarrow l^+\nu$. 

\section{Determination of CKM matrix elements $|V_{\rm cd}|$ and $|V_{\rm cs}|$}
    The CKM elements $|V_{\rm cd}|$ and $|V_{\rm cs}|$ can be extracted from
the leptonic decay branching fractions of the $D^+$ and $D_s^+$ mesons. In this section, we discuss
the determinations of the $|V_{\rm cd}|$ and $|V_{\rm cs}|$ with the measured branching fractions
for these leptonic decays and test of the unitary of the CKM matrix. 

\subsection{Measurements of $|V_{\rm cd}|$ and $|V_{\rm cs}|$ with $D_{(s)}^+ \rightarrow l^+\nu$}

    The BES-III determined the CKM matrix element $|V_{\rm cd}|$ with the measured branching fraction for
$D^+ \rightarrow \mu^+\nu$ decay.
Inserting the branching fraction measured at the BES-III, the mass of the muon,
the mass of the $D^+$ meson, the decay constant $f_{D^+}=207\pm 4$ MeV from LQCD~\cite{lqcd_HPQCD_UKQCD},
$G_F$ and the lifetime of the $D^+$ meson~\cite{pdg2010}
into Eq.(\ref{eq01}) yields
$$|V_{\rm cd}| = 0.2218 \pm 0.0062 \pm 0.0047~~({\rm BESIII~Preliminary}),$$
\noindent
where the first error is statistical and the second systematic arising
mainly from the uncertainties in
the measured branching fraction ($1.7\%$),
$f_{D^+}$ ($1.93\%$),
and the lifetime of the $D^+$ meson ($0.7\%$)~\cite{pdg2010}.
The total systematic error is $2.1\%$.
Table~\ref{comparison_measured_vcd} lists
the comparison of the measured magnitude of $V_{\rm cd}$ from different experiments.

\begin{table}[htbp]
\caption{Comparison of the measured $|V_{\rm cd}|$.}
\label{comparison_measured_vcd}
\begin{center}
\begin{tabular}{lr} \hline\hline
Experiment                                                            &   $|V_{\rm cd}|$    \\  \hline
PDG10 (Charm decays)~\cite{pdg2010}                                   &   $0.229 \pm 0.006 \pm 0.024$ \\
PDG10 ($\nu$ and $\bar \nu$ interaction)~\cite{pdg2010}               &   $0.230 \pm 0.011$           \\
CLEO-c ($D\rightarrow \pi e^+\nu_e$)~\cite{cleo-c_Phys_Rev_D80_032005_y2009}
                                                                      & $0.234 \pm 0.007 \pm 0.002 \pm 0.025$ \\
BES-III ($D^+ \rightarrow \mu^+\nu_{\mu}$)                            & $0.222 \pm 0.006 \pm 0.005$ \\
\hline \hline
\end{tabular}
\end{center}
\end{table}

\begin{figure}
\centerline{
\includegraphics[width=9.0cm,height=9.0cm]
{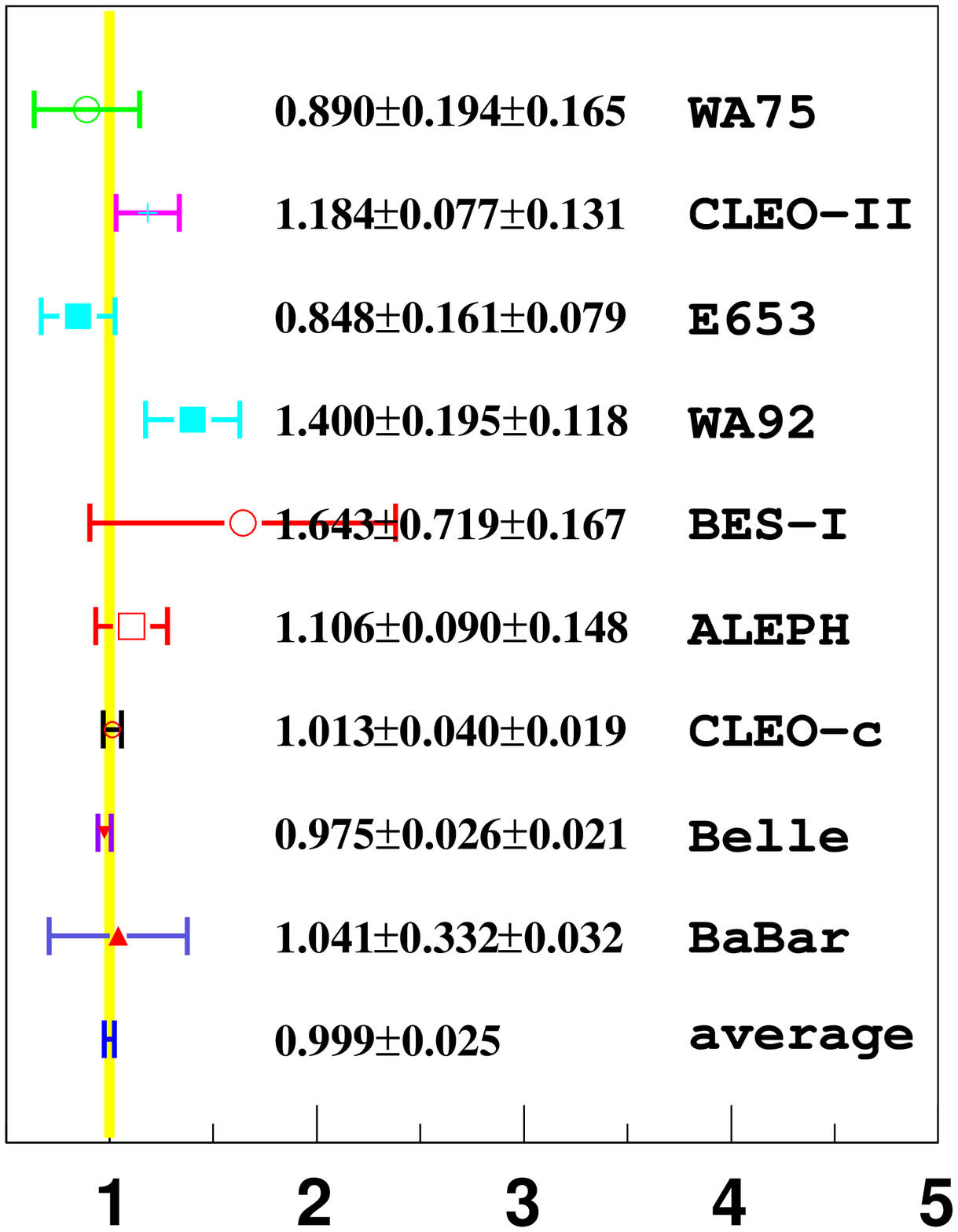}
\put(-140.0,12.0){\large{$|V_{\rm cs}|$ }}
\put(-50.0,220.0){(a)}
\put(-160.0,235.0){$D_s^+\rightarrow \mu^+\nu$}
\includegraphics[width=9.0cm,height=9.0cm]
{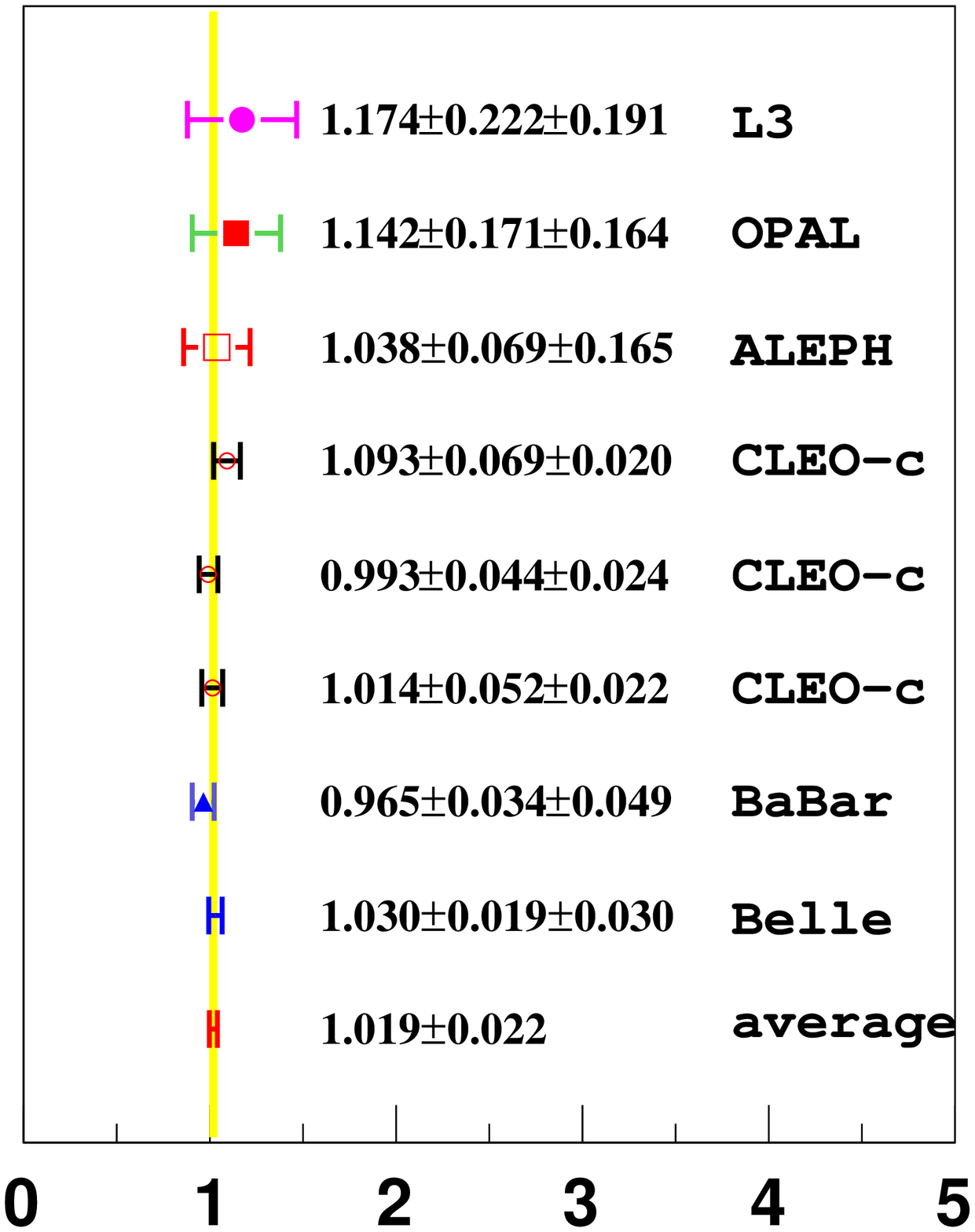}
\put(-140.0,12.0){\large{$|V_{\rm cs}|$}}
\put(-50.0,220.0){(b)}
\put(-165.0,235.0){$D_s^+\rightarrow \tau^+\nu$}
        }
\caption{Comparison of the measured $|V_{\rm cs}|$ obtained from (a) $D_s^+ \rightarrow \mu^+\nu$ decay
and (b) $D_s^+ \rightarrow \tau^+\nu$ decay.}
\label{fig9}
\end{figure}

\begin{figure}
\centerline{
\includegraphics[width=9.0cm,height=9.0cm]
{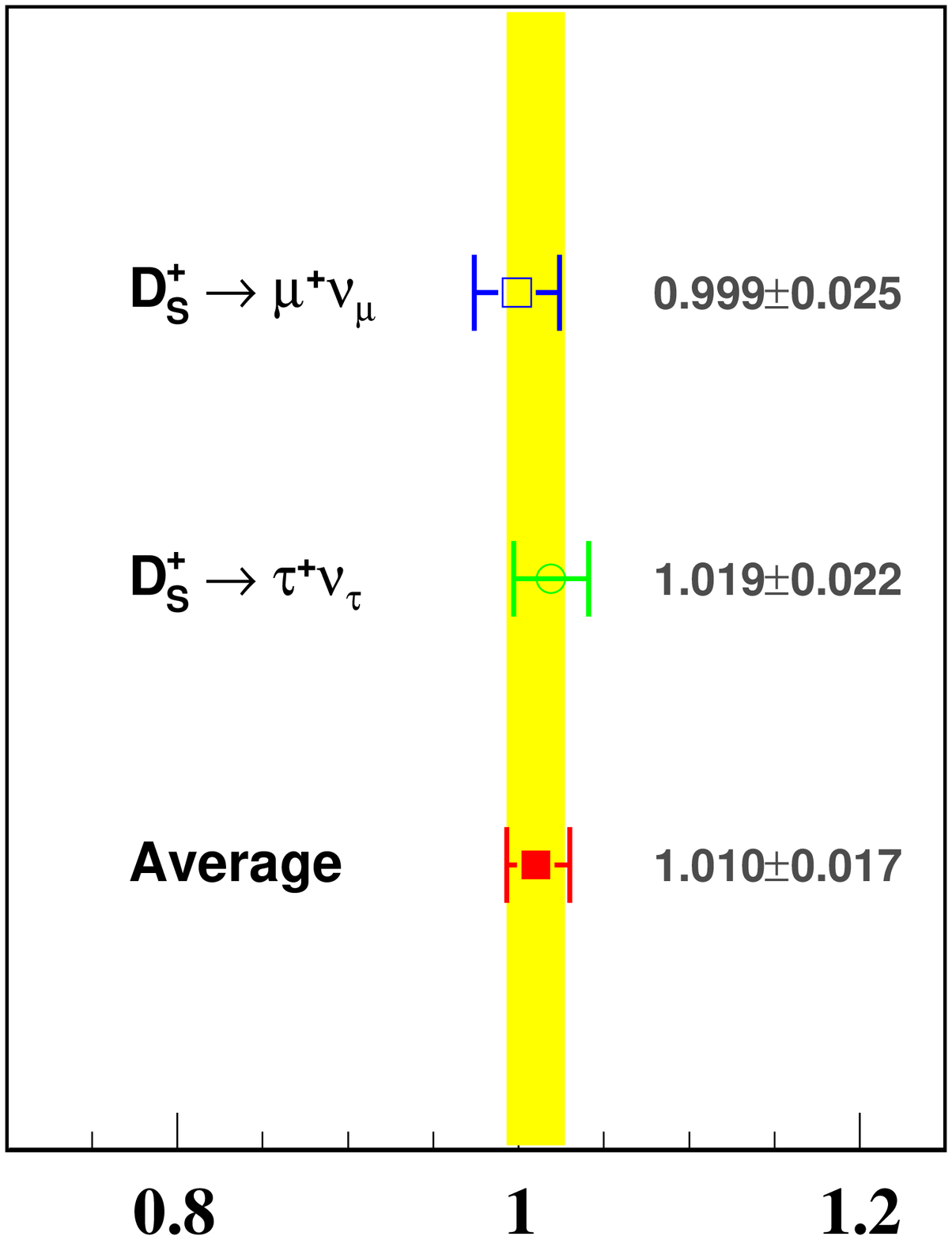}
\put(-140.0,12.0){\large{$|V_{\rm cs}|$ }}
\put(-50.0,220.0){(a)}
\includegraphics[width=9.0cm,height=9.0cm]
{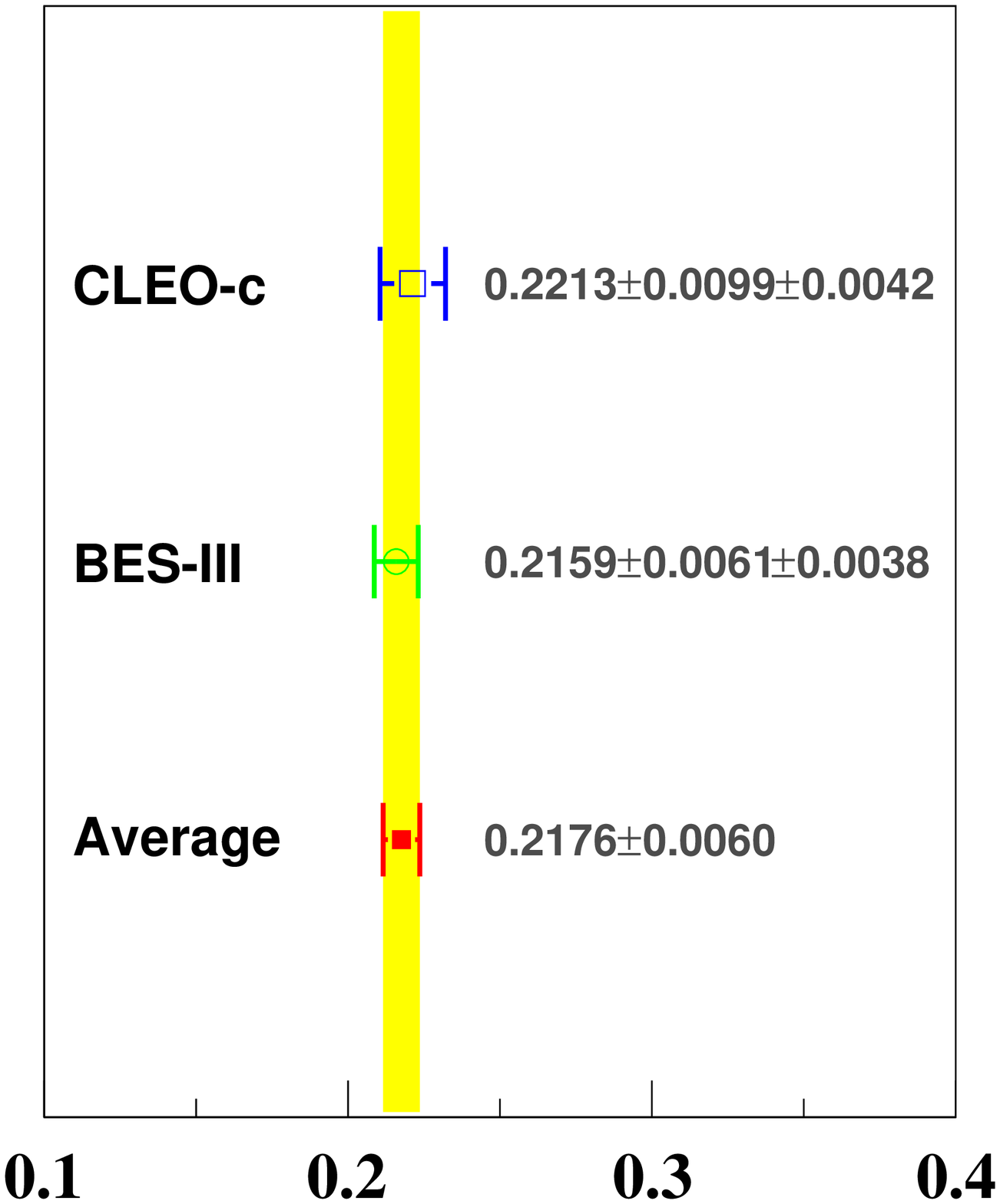}
\put(-140.0,12.0){\large{$|V_{\rm cd}|$}}
\put(-50.0,220.0){(b)}
        }
\caption{Comparison of the measured (a) $|V_{\rm cs}|$ obtained from $D_s^+ \rightarrow \mu^+\nu$
and $D_s^+ \rightarrow \tau^+\nu$ decays, and (b) $|V_{\rm cd}|$ obtained from $D^+ \rightarrow \mu^+\nu$ decay.}
\label{fig10}
\end{figure}

   With the $f_{D_s^+}=(247.5\pm 2.2)$ MeV which is the weighted average of the predicted decay constants 
with theories based on QCD,
we calculate the CKM matrix element $|V_{\rm cs}|$ with the branching fractions for
$D_{S}^+ \rightarrow \mu^+\nu$ and $D_{S}^+ \rightarrow \tau^+\nu$ decays measured at different experiments. 
Figure~\ref{fig9} (a) and (b) show comparison of the $|V_{\rm cs}|$ determined with decay branching fractions
for $D_s^+ \rightarrow \mu^+\nu$ and $D_s^+ \rightarrow \tau^+\nu$ measured at different experiment, respectively.
Figure~\ref{fig10} (a)  shows the $|V_{\rm cs}|$ determined with decay branching fractions
for $D_s^+ \rightarrow \mu^+\nu$ and $D_s^+ \rightarrow \tau^+\nu$ as well as the
average of the $|V_{\rm cs}|$. The world average of $|V_{\rm cs}|$ is $1.010 \pm 0.017$.

   With the $f_{D^+}=(212.7\pm 3.2)$ MeV which is the weighted average of the predicted decay constants 
with theories based on QCD,
we calculate the CKM matrix element $|V_{\rm cd}|$ with the branching fractions for
$D^+ \rightarrow \mu^+\nu$ decay measured at the BES-III and the CLEO-c. 
Figure~\ref{fig10} (b) shows comparison of the $|V_{\rm cd}|$ determined with decay branching fractions
for $D_s^+ \rightarrow \mu^+\nu$ measured at different experiment as well as the average of the $|V_{\rm cd}|$.
The world average of $|V_{\rm cd}|$ is $0.2176\pm0.0060$.

    Comparing the CKM matrix elements of $|V_{\rm cd}|$ and  $|V_{\rm cs}|$ obtained by analyzing 
the $D^+$ and $D_{s}^+$
leptonic decays with those obtained by analyzing $D$ meson semileptonic decays can provides 
some useful information about the New Physics beyond the Standard Model. 
If no nonstandard leptonic decay of the $D^+_{s}$ meson, the values of $|V_{\rm cd(s)}|$ 
determined from the $D^+_{s}$ leptonic decay branching fraction and determined from the $D_{(s)}$ meson
semileptonic decay branching fractions should be the same.
From the CKMfitter~\cite{pdg2010} one obtains $|V_{\rm cs}|_{\rm CKMfitter}=(0.97345^{+0.00015}_{-0.00016})$ 
and $|V_{\rm cd}|_{\rm CKMfitter}=(0.2252\pm 0.0007)$. While from the leptonic decays of the $D_s^+$ and $D^+$ mesons
we obtain $|V_{\rm cs}|_{D_s^+\rightarrow l^+\nu}=(1.010\pm 0.017)$
and $|V_{\rm cd}|_{D^+\rightarrow l^+\nu}=(0.218\pm 0.006)$.
However, from $D$ meson semileptonic decays, the CLEO-c measured
$|V_{\rm cs}|_{\rm {D~semileptonic~decays}}=0.985\pm0.009\pm0.006\pm0.103$~\cite{cleo-c_Phys_Rev_D80_032005_y2009} and
$|V_{\rm cd}|_{\rm {D~semileptonic~decays}}=0.234\pm0.007\pm0.002\pm0.025$~\cite{cleo-c_Phys_Rev_D80_032005_y2009}.
From the values of these $|V_{\rm cd}|$ we found that the $|V_{\rm cd}|$ determined from $D^+$ leptonic decays
is consistent within the error with these either determined from the $D$ meson semileptonic decays
or determined from the CKMfitter~\cite{pdg2010}.
However, comparing the values of $|V_{\rm cs}|$ we found that the $|V_{\rm cs}|$ 
determined from the CLMfiter 
is consistent within error with the one determined from $D$ semileptonic decays,
but the value of the $|V_{\rm cs}|$ determined from $D_s^+$ leptonic decays
is $2.2\sigma$ larger than the $|V_{\rm cs}|_{\rm CKMfitter}$.
This $2.2\sigma$ deviation of $|V_{\rm cs}|_{D_s^+\rightarrow l^+\nu}$
from the $|V_{\rm cs}|_{\rm CKMfitter}$
may indicate that there are some New Physics effects which enhance the $D_s^+$ leptonic decays.

    To make more precision comparison of these CKM matrix elements, we need to reduce the errors of
the measured decay branching fractions of $D_s^+$ and $D^+$ leptonic and semileptonic decays
as well as reduce the errors of the measured decay branching fractions of the $D^0$ semileptonic decays.
These could well be done at the currently running BES-III experiment and at the BELLE and BaBar experiments.
At present, two analysis working groups (IHEP and CMU groups)~\cite{liucl_talk} 
in the BES-III collaboration have been working on extracting the
$|V_{\rm cs}|$ and $|V_{\rm cd}|$ as well as other physical quantities 
from $D$ meson semileptonic decays. 
Based on these two working group analysis of $D^0$ semileptonic decays with a portion of data taken at 3.773 GeV, 
the BES-III collaboration report preliminary results on measurements of $|V_{\rm cs}|f_+^K(0)$ 
and $|V_{\rm cd}|f_+^{\pi}(0)$~\cite{liucl_talk}, 
where $f_+^K(0)$ and $f_+^{\pi}(0)$ are the form factors $D$ semileptonic decays.
More precision measurements of $|V_{\rm cs}|$ and $|V_{\rm cd}|$  
would be important physics results
for precise test of the SM and search for New Physics.

\section{Conclusion and Outlook}

    Since the first attempt to search for the $D^+$ leptonic decay,
although no signal event was found for this decay performed at the MARK-III experiment in 1988,
many experiments have been making great efforts to search for and study the $D^+$ and $D_s^+$ leptonic decays.
After more than 25 year studies of $D^+$ and $D_s^+$ leptonic decays, more than 530
$D^+ \rightarrow \mu^+\nu$ and about $4\times 10^3$ $D_s^+ \rightarrow l^+\nu$ decay events have been found.
One begins to precisely study the hadronic vertex of the $D^+$ and $D_s^+$ meson decays and precisely test the
LQCD calculations of the decay constants $f_{D^+}$ and $f_{D_s^+}$.
At the Charm2012 conference, the BES-III collaboration report
the most precise results for measurements of the decay branching fraction,
decay constant and $|V_{\rm cd}|$ in the world, 
which are $B(D^+ \rightarrow \mu^+\nu)=(3.74 \pm 0.21 \pm 0.06)\times 10^{-4}$,
$f_{D^+}=(203.9\pm 5.7 \pm 2.0)$ MeV, and 
$|V_{\rm cd}|=(0.222 \pm 0.006 \pm 0.005)$.
The most precise measurements of
$B(D^+_s \rightarrow l^+\nu)$ and $f_{D_s^+}$ are from the BELLE experiment. The BELLE
results are
$B(D_s^+ \rightarrow \mu^+\nu)=(0.528 \pm 0.028 \pm 0.019)\%$,
$B(D_s^+ \rightarrow \tau^+\nu)=(5.70 \pm 0.21^{+0.31}_{-0.30})\%$,
and $f_{D_s^+}=(255.0\pm 4.2 \pm 5.0)$ MeV.

The world average of decay constants are 
$f_{D^+}=(205.3\pm 5.1)$ MeV and
$f_{D_s^+}=(256.8\pm 4.1)$ MeV.
The two decay constants
yield the world average of the ratio
of $f_{D_s^+}/f_{D^+} = 1.251\pm 0.037$,
which is about 2.5$\sigma$ larger than $f_{D_s^+}/f_{D^+} = 1.156 \pm 0.007$
predicted with theories based on QCD.
By comparing the $|V_{\rm cs}|$ and $|V_{\rm cd}|$ determined 
from the $D_s^+$ and $D^+$ meson leptonic decays,
determined from the CKMfitter~\cite{pdg2010},
and determined from the $D$ meson semileptonic decays,
we also found that
the $|V_{\rm cs}|$ determined from $D_s^+$ leptonic decays
is $2.2\sigma$ larger than the $|V_{\rm cs}|_{\rm CKMfitter}$.
This $2.2\sigma$ deviation of $|V_{\rm cs}|_{D_s^+\rightarrow l^+\nu}$ from the $|V_{\rm cs}|_{\rm CKMfitter}$
may indicate that there are some New Physics effects which enhance the $D_s^+$ leptonic decays.

    The measured $D^+$ and $D^+_s$ meson decay constants $f_{D^+}$ and $f_{D_s^+}$ can 
be used to test LQCD calculations of
the decay constants. At present the uncertainties of the measured decay constants
are almost the same as the uncertainties of the LQCD calculations of the decay constants.
To more precisely test the LQCD calculations of the decay constants,
we still need more data to be collected at 3.773 GeV, at energy near $D_s^+ D_s^-$ meson pair 
production energy threshold, and at the higher energy of about 10.6 GeV.
These data taking will be performed at the BES-III and other B factory experiments in the future.
The verified LQCD calculation help extract $|V_{\rm td}|$ and
$|V_{\rm ts}|$ from $B\bar B$ mixing experiments.
These help more precisely test the SM and search for New Physics beyond the SM.

\vspace{0.2cm}

\section*{Acknowledgement}
I gratefully acknowledge my colleague, Mr. Y. Fang, Dr. L.L. Jiang and Dr. H.L. Ma for helping me prepare
some plots and check some figures in the article. 
I would like to thank Prof. M. Mandelkern, Prof. S. Olsen, Prof. D. Kirkby and Dr.
H. Muramatsu for letting me know some information about leptonic decays of $D^+$ and $D_s^+$ mesons
at some $e^+e^-$ experiments.
I wish to thank Prof. R. Briere for letting me know CLEO updated analysis of $D_s^+ \to \mu^+\nu$ decays.
This work is partly supported by National Natural Science Foundation of China (10935007) 
and National Key Basic Research Program (973 by MOST) (2009CB825200XX).


\vspace{0.0cm}

\begin{thebibliography}{9}

\bibitem{quarks_and_leptons} Francis Halzen, Alan D. Martin, 
                             Quarks \& Leptons (John Wiley \& Sons, New York, Chichester, Brisbane, Toronto, Singapore, 1984).

\bibitem{pdg2010} K. Nakamura $et~al.$ J. Phys. G {\bf 37}, 075021 (2010).
\bibitem{QCD-inspired_potential_model} S. Capstick and S. Godfrey, Phys. Rev. D{\bf 41}, 2856 (1990);
                                       P. Colangelo, G. Nardulli and M. Pietroni, Phys. Rev. D{\bf 43}, 3002 (1991).
\bibitem{QCD_sum_rules} M. Neubert, Phys. Rev. D{\bf 45}, 2451 (1992);
                        E. Bagan $et ~al.$, Phys. Lett. B{\bf 278}, 457 (1992);
                        K. Schilcher and Y.L. Wu, Z. Phys. C{\bf 54}, 163 (1992);
                        C.A. Dominguez and N. Paver, Phys. Lett. B{\bf 197}, 423 (1987); B{\bf 199}, 596(E) (1987);
                        S. Narison, Phys. Lett. B{\bf 198}, 104 (1987);
                        T.M. Aliev and V.L. Eletskii, Sov. J. Nucl. Phys. 38, 936 (1983).
\bibitem{simulation_of_QCD_on_lattice} A. Abada $et ~al.$, Nucl. Phys. B{\bf 376}, 172 (1992);
                                       M.B. Gavela $et ~al.$, Phys. Lett. B{\bf 206},113 (1988);
                                       C. Alexandrou $et ~al.$, Phys. Lett. B{\bf 256}, 60 (1991);
                                       C. Bernard $et ~al.$, Phys. Rev. DB{\bf 38}, 3540 (1988);
                                       T. A. DeGrand and R. D. Loft, Phys. Rev. DB{\bf 38}, 954 (1988). 
\bibitem{non-perturbative_methods} Y. A. Simonov, Z. Phys. C{\bf 53}, 419 (1992);
                                   R. R. Mendel and H. D. Trottier, Phys. Lett. B{\bf 231}, 312 (1989);
                                   D. Izatt, D. DeTar and M. Stenphenson, Nucl. Phys. B{\bf 199}, 269 (1982).
\bibitem{lqcd_HPQCD_UKQCD} E. Follana $et ~al.$, (HPQCD and UKQCD Collaborations),
                           Phys. Rev. Lett. {\bf 100},062002 (2008).

\bibitem{Dobrescu_and_Kronfeld} B.A. Dobrescu and A.S. Kronfeld, Phys. Rev. Lett. {\bf 100}, 241802 (2008).

\bibitem{Kundu_and_Nandi} A. Kundu and S. Nandi, Phys. Rev. D {\bf 78}, 015009 (2008).  

\bibitem{mark-iii_fD} J. Adler $et ~al.$ (The MARK III Collaboration), Phys. Rev. Lett. {\bf 60}, 1375 (1998). 

\bibitem{bes-i_fD} J.Z. Bai $et ~al.$ (BES Collaboration), Phys. Lett. B{\bf 429}, 188 (1998).

\bibitem{bes-ii_fD} G. Rong (for BES Collaboration), 
                    Proceeding of the XXXIXth RENCONTRES DE MORIOND, March 21--28, 2004, edited by J. Tran Thanh Van; 
                    M. Ablikim $et ~al.$ (BES Collaboration), Phys. Lett. B{\bf 610}, 183 (2005).


\bibitem{cleo-c_fD_2004} G. Bonvicini $et ~al.$ (CLEO Collaboration), Phys. Rev. D {\bf 70}, 112004 (2004).

\bibitem{cleo-c_fD_2005} M. Artuso $et ~al.$ (CLEO Collaboration), Phys. Rev. Lett. {\bf 95}, 251801 (2005).

\bibitem{cleo-c_fD_2008} B.I. Eisenstein $et ~al.$ (CLEO Collaboration), Phys. Rev. D {\bf 78}, 052003 (2008).

\bibitem{bes3} M. Ablikim $et ~ al.$ (BESIII Collaboration), Nucl. Instrum. Methods Phys. Res., Sect. A {\bf A 614}, 345 (2010)
\bibitem{bepc2} J. Z. Bai $et ~ al.$ (BES Collaboration), Nucl. Instrum. Methods Phys. Res.  {\bf A 458}, 627 (2001).
\bibitem{WA75} S. Aoki $et ~al.$ (WA75 Collaboration), Prog. Theor. Phys. {\bf 89}, 131 (1993).
\bibitem{E653} K. Kodama $et ~al.$ (E653 Collaboration), Phys. Lett. B{\bf 382}, 299 (1996).
\bibitem{WA92} Y. Alexandrov $et ~al.$ (BEATRICE Collaboration), Phys. Lett. B{\bf 478}, 31 (2000).

\bibitem{bes-i_fDs} J.Z. Bai $et ~al.$ (BES Collaboration), Phys. Rev. Lett.{\bf 74}, 4599 (1995).


\bibitem{cleo-c_fDs_2009a} J.P. Alexander  $et ~al.$ (CLEO Collaboration), Phys. Rev. D {\bf 79}, 052001 (2009).
\bibitem{cleo-c_fDs_2009b} P.U.E. Onyisi  $et ~al.$ (CLEO Collaboration), Phys. Rev. D {\bf 79}, 052002 (2009).
\bibitem{cleo-c_fDs_2009c} P. Naik  $et ~al.$ (CLEO Collaboration), Phys. Rev. D {\bf 80}, 112004 (2009).

\bibitem{cleo_y1994_fDs} M. Chada $et ~al.$ (CLEO Collaboration), Phys. Rev. D {\bf 58}, 032002 (1998).
\bibitem{belle_y2008_fDs} L. Widhalm $et ~al.$ (The BELLE Collaboration), Phys. Rev. Lett. {\bf 100}, 241801 (2008).
\bibitem{belle_charm2012_fDs} Anze Zupanc (for BELLE Collaboration), ''New Belle result on $f_{D_s}$ + experimental status of 
                 $f_{D_s}$ and $f_D$'', The 5th International Workshop on Charm Physics, Honolulu, Hawaii, May, 2012.
\bibitem{babar_y2008_fDs} P. del Amo Sanchez $et ~al.$ (BaBar Collaboration), Phys. Rev. D {\bf 82}, 091103(R) (2010).
\bibitem{l3_fDs} M. Acciarri  $et ~al.$ (L3 Collaboration), Phys. Lett. B {\bf 396}, 327 (1997)
\bibitem{opal_fDs} G. Abbiendi $et ~al.$ (OPAL Collaboration), Phys. Lett. B {\bf 516}, 236 (2001)
\bibitem{aleph_fDs}  R. Barate $et ~al.$ (ALEPH Collaboration), Phys. Lett. B {\bf 528}, 1 (2002)

\bibitem{lqcd_Fermilab_MILC} C. Aubin, $et ~al.$, (FNAL Lattice, HPQCD, MILC)
                             Phys. Rev. Lett. {\bf 95}, 122002 (2005).
\bibitem{Khan_pkb652_p150_y2007} A. Ali Khan $et ~al.$ (QCDSF Collaboration), Phys. Lett. B {\bf 652}, 150 (2009).

\bibitem{Chiu_plb642_p31_2005} T.W. Chiu $et ~al.$, Phys. Lett. B {\bf 624}, 31 (2005).

\bibitem{Lellouch_prd64_p094501_y2001} L. Lellouch and C.-J David Lin (UKQCD), Phys. Rev. D {\bf 64}, 094501 (2001).

\bibitem{Becirevic_prd60_p074501_1999} D. Becirevic $et ~al.$, Phys. Rev. D {\bf 60}, 074501 (1999).

\bibitem{Bordes_j_high_energy_phiscs11_p014_y2005} J. Bordes $et ~al.$, J. High Energy Phys. 11, 104 (2005).

\bibitem{Narison_hep-ph-0202200} S. Narison, arXiv:hep-ph/0202200.

\bibitem{Badalian_prd75_p116001_y2007} A.M. Badalian $et ~al.$, Phys. Rev. D {\bf 75}, 116001 (2007).

\bibitem{Penin_prd65_p054005_y2002} A. Penin and M. Steinhauser, Phys. Rev. D {\bf 65}, 054006 (2002).

\bibitem{Amundson_prd47_p3059_y1993} J. Amundson $et ~al.$, Phys. Rev. D {\bf 47}, 3059 (1993).

\bibitem{cleo-c_Phys_Rev_D80_032005_y2009}  CLEO Collaboration, D. Besson $et ~al.$, Phys. Rev. D {\bf 80}, 032005(2009).

\bibitem{liucl_talk} C.L. Liu (for BES-III Collaboration), ''Recent Results of $D$ Semi-leptonic Decays'', 
                     The 5th International Workshop on Charm Physics, Honolulu, Hawaii, May, 2012.

\end{thebibliography}
\end{document}